\documentclass[fleqn,usenatbib]{mnras}

\usepackage{newtxtext,newtxmath}
\usepackage[T1]{fontenc}
\usepackage{ae,aecompl}

\usepackage{graphicx}	
\usepackage{amsmath}	
\usepackage{amssymb}	
\usepackage{hyperref}

\newcommand{\kpc}{\mbox{kpc}}
\newcommand{\mpc}{\mbox{Mpc}}
\newcommand{\Mpch}{\mbox{$h^{-1}$ Mpc}}
\newcommand{\Msun}{\mbox{$M_\odot$}}

\title[DM halos mass \& velocity functions]{Accurate mass and velocity functions of dark matter halos}

\author[Comparat et al.]{
Johan Comparat$^{1,2,3}$\thanks{comparat@mpe.mpg.de}, 
Francisco Prada$^{4}$, 
Gustavo Yepes$^{2}$, 
Anatoly Klypin$^{5}$
\\
$^1$Instituto de F\'{\i}sica Te\'orica UAM/CSIC, 28049 Madrid, Spain\\
$^2$Departamento de F\'{\i}sica Te\'orica, Universidad Aut\'onoma de Madrid, 28049 Madrid, Spain\\
$^3$Max-Planck-Institut f\"{u}r extraterrestrische Physik (MPE), Giessenbachstrasse 1, D-85748 Garching bei München, Germany\\
$^4$Instituto de Astrof\'{\i}sica de Andaluc\'{\i}a (CSIC), Glorieta de la Astronom\'{\i}a, E-18080 Granada, Spain \\
$^5$Astronomy Department, New Mexico State University, Las Cruces, NM, USA
}

\date{Accepted XXX. Received 02.2017; in original form ZZZ}

\pubyear{2017}

\begin{document}

\let\oldpageref\pageref
\renewcommand{\pageref}{\oldpageref*}

\maketitle

\begin{abstract}
$N$-body cosmological simulations are an essential tool to understand the observed distribution of galaxies. 
We use the MultiDark simulation suite, run with the Planck cosmological parameters, to revisit the mass and velocity functions. 
At redshift $z=0$, the simulations cover four orders of magnitude in halo mass from $\sim10^{11}M_\odot$ with 8,783,874 distinct halos and 532,533 subhalos. The total volume used is $\sim$515 Gpc$^3$, more than 8 times larger than in previous studies. 
We measure and model the halo mass function, its covariance matrix w.r.t halo mass and the large scale halo bias. 
With the formalism of the excursion-set mass function, we explicit the tight interconnection between the covariance matrix, bias and halo mass function. 
We obtain a very accurate ($<2\%$ level) model of the distinct halo mass function. We also model the subhalo mass function and its relation to the distinct halo mass function. 
The set of models obtained provides a complete and precise framework for the description of halos in the concordance Planck cosmology.
Finally, we provide precise analytical fits of the $V_{max}$ maximum velocity function up to redshift $z<2.3$ to push for the development of halo occupation distribution using $V_{max}$. 
The data and the analysis code are made publicly available in the \textit{Skies and Universes} database.
\end{abstract}



\section{Introduction}
N-body cosmological simulations are essential tools to understand the observed distribution of galaxies. 
In the last decades, development of numerical codes \citep{Teyssier2002,Springel2005,Springel2010,Klypin2011,Habib2016} and the access to powerful supercomputers enabled the computation of high resolution cosmological simulations over large volumes \textit{e.g.} MultiDark \citep[MD hereafter,][]{Prada2012}; DarkSkies \citep[DS hereafter,][]{Skillman2014}. 
Both simulations were run in the paradigm of the flat Lambda Cold Dark Matter cosmology \citep[$\Lambda$CDM,][]{Planck2014}. From MD emerged the most precise description to date of the dark matter halo \citep{Klypin2016}. 
While finding and describing the halos formed by the dark matter is now well understood \citep{Behroozi2013,Knebe2013,Avila2014}, connecting galaxies to halos is a proven complicated subject. 
There are three main streams of galaxy assignment in simulations, we order them by decreasing computational needs and accuracy: 
\textit{(i)} hydrodynamical simulations \citep[HYDRO,][]{Cen1993,Springel2003}, 
\textit{(ii)} semi-analytical models of galaxy formation \citep[SAMS,][]{Cole2000,Baugh2006}, 
\textit{(iii)} halo occupation distribution or subhalo abundance matching \citep[HOD, SHAM,][respectively]{Cooray2002,Conroy2006}. 
The existing methods will hopefully converge in the coming years \citep{Knebe2015,Elahi2016,Guo2016}. 

The current and future cosmological galaxy and quasar surveys, e.g. BOSS, eBOSS, DES, DESI, 4MOST, Euclid, will cover gigantic volumes up to redshift 3.5 \citep{Dawson2013,2016AJ....151...44D,Abbott_2005,DESI2016,2011arXiv1110.3193L}. 
These volumes are too large to be entirely simulated with hydrodynamics. There is thus a need to improve the predictive power of the SAMS and HOD to the level of the expected 2-point function measurements, i.e. around the percent level. 
This challenge needs to be handled from both, the hydrodynamical simulation point of view \citep{Chaves-Montero2015,Sawala2015} and from the DM-only simulation perspective \citep{RodriguezTorres2015,favole_2015_elg,Carretero2015} to eventually join in an optimal semi-analytical model \citep{Knebe2015}. Lastly, \citet{Castro2016} argued that with such surveys, one would constrain directly the parameters of the mass function to the level that it is estimated in $N$-body simulations, enhancing again the need of a precise model for the halo mass function (HMF).

From the DM-only simulation perspective, the most fundamental statistic is the halo mass function. Observational probes, such as weak lensing, galaxy clustering or galaxy clusters, also rely on the knowledge of the halo mass function. 
The mass function denotes, at a given redshift, the fraction of mass contained in collapsed halos with a mass in the interval $M$ and $M+dM$. 
It was studied theoretically and numerically in various simulations and different cosmologies \citep{PressSchechter1974,ShethTormen1999,Sheth2001,Sheth2002,Jenkins2001,Springel2005b,Warren2006,Tinker2008,Bhattacharya2011,Angulo2012,Watson2013,Despali2016}. 

The theoretical formalism to describe the number density of halos was initiated by \citet{PressSchechter1974}. Its latest formulation by \citet{Sheth2001,ShethTormen1999} includes the ellipsoidal collapse instead of spherical collapse. Heuristically, it corresponds to a diffusion across a `moving' or across a mass-dependent boundary. The excursion set formalism of the mass function constitutes today a good description of what is measured in $N$-body simulations. More precise predictions are actively being sought and eventually we might converge towards an ultimate universal mass function. The variety of existing and tested functional forms of the mass function are discussed and compared in \citet{Murray2013}. The description of the errors on the HMF is slightly less discussed subject. Nevertheless, \citet{HU2003,Bhattacharya2011} provided a solid background, used in this study, to model errors on the HMF and the large-scale halo bias. 

Numerically, the HMF was extensively studied with a cosmology-independent (universal) model. The most recent measurements on $N$-body simulations enabled models to predict any HMF to about 10\% accuracy; see \citet{Despali2016}. It is to date the latest HMF measurements in the Planck cosmology. We feel though, the lack of a percent-level-accurate model for the HMF in the Planck cosmology.

The recent measurements of the cosmic microwave background indicate a significantly higher matter content than suggested by previous observations \citep[WMAP,][]{WMAP72011}. 
And the matter content of the Universe is a parameter that strongly influences the HMF. 
We think it is thus necessary to revisit the parametrization of the mass function and understand to what accuracy the mass function is known in our best cosmological model. 
Previous works could not assess thoroughly the uncertainties on the measurement of the mass function due to the limited amount of $N$-body realizations available. 
With the MD and DS simulations, extracting covariance matrices becomes possible. 

In this paper, we explore and model the HMF and its covariance matrix. We describe the model in Section \ref{sec:model}. In Section \ref{sec:data}, we describe the simulations used and we estimate the halo mass function, its covariance and the large scale halo bias.
The HMF results are presented in Section \ref{sec:resultMvirFunction}. Finally, in Appendix \ref{sec:vmax:fun} we parametrize the redshift evolution of the distinct and satellite halo velocity function.

\subsection*{Data base}
All the data and the results are available through the \textit{Skies and Universes} database\footnote{\href{http://projects.ift.uam-csic.es/skies-universes/}{projects.ift.uam-csic.es/skies-universes/}}. The code is made public via GitHub\footnote{\href{https://github.com/JohanComparat/nbody-npt-functions}{github.com/JohanComparat/nbody-npt-functions}}.

\section{Model}
\label{sec:model}

\subsection{Halo mass function}
The formalism to describe the number density of halos 
was initiated by \citet{PressSchechter1974}. 
They assumed that the fraction of mass in halos of mass greater than $M$ at a time $t$, $F(>M, t)$, was equal to twice the probability, $\mathcal{P}$, for the smoothed density field, $\delta_s$, to overcome the critical threshold for spherical collapse, $\delta_c$ i.e.
\begin{equation}
F(>M, t) = 2 \mathcal{P}(\delta_s(t) > \delta_c(t)).
\end{equation}
Assuming that $\delta_s$ is a Gaussian random field, they related the number density of halos to $F$
\begin{equation}
n(M,t) dM = \frac{\bar{\rho}}{M} \frac{\partial F(>M, t)}{\partial M} dM.
\end{equation}
The mass function depends on redshift and on halo mass. 
Rather than mass, it is physically more relevant to use the root mean square ($RMS$) fluctuations of the linear density density field smoothed with a filter encompassing this mass 
\begin{equation}
\sigma^2(M,t) = 4\pi^2 \int_0^\infty P(k,t) W^2(k,M) k^2 dk,
\label{eqn:m:sigma}
\end{equation}
where $P(k)$ is the linear power spectrum and $W$ a top-hat filter. \newline
Assuming that the initial Gaussian random density fluctuation field evolves and crosses via a random walk the spherical collapse barrier, these equations determine the number of regions in the simulation that underwent collapse at a given time
\begin{equation}
n(\sigma, t) dM =  f_{PS}(\sigma) \frac{\bar{\rho}}{M^2} \frac{d\ln \sigma}{d\ln M} dM,
\end{equation}
where the function $f$, called the multiplicity function has the following expression
\begin{equation}
f_{PS}(\sigma) = \sqrt{\frac{2}{\pi}} \frac{\delta_c}{\sigma} \exp{\left[- \frac{\delta^2_c }{ 2\sigma^2}\right]}.
\end{equation}
`PS' stand for `Press Schechter'. In other words, it is the fraction of mass associated with halos in a unit range of $d\ln\sigma$. 
Because the threshold $\delta_c$ increases with time, smaller halos are formed first and then the larger ones (hierarchical clustering).

This model was revised using excursion set theory by \citet{Bond1991}. They argued that $\sigma$ diffuses across the spherical collapse boundary or barrier, instead of crossing it via a random walk. This lead to a new multiplicity function
\begin{equation}
f_{EPS}(\sigma) = f_{PS}(\sigma) / (2\sqrt{\sigma})
\end{equation}
where `EPS' stand for Extended-Press-Schechter.

\citet{ShethTormen1999,Sheth2001} later explored the ellipsoidal collapse to replace the assumption of spherical collapse. Heuristically, it corresponds to a diffusion across a `moving' barrier (or across a $\sigma$ dependent boundary). They found the following multiplicity function $f_{ST}$, 
\begin{equation}
f_{ST}(\sigma, A, a, p) = A \sqrt{\frac{2}{\pi}}\left[1+\left(\frac{\sigma^2}{a \delta^2_c}\right)^p\right] \left(\frac{\sqrt{a}\delta_c}{\sigma}\right) \exp\left[ -\frac{a}{2} \frac{\delta^2_c}{\sigma^2} \right],
\label{eqn:mf:model:st02}
\end{equation}
where `ST' stands for `Sheth and Tormen'. 
It constitutes a further improvement compared to $f_{EPS}$. 

The latter multiplicity function describes well the $\Lambda$CDM distinct halo mass function with the parameters (A, a, p)=(0.3222, 0.707, 0.3). 
These parameters were measured again by \citet{Despali2016} in the latest Planck-cosmology paradig. They found (A, a, p)=(0.333, 0.794, 0.247). 
It remains a statistical scatter of the simulated data around this model of order of 5 to 7\% at the high mass end. 
More precise predictions are actively being sought \citep[e.g.][]{Pace2014,Reischke2016MNRAS,Popolo2017}. Eventually we will converge towards an ultimate physical model for the halo mass function.

Aside from the physical model of the mass function, exist a variety of functional forms created to best fit the mass function as measured in $N$-body simulations; see \citet{Murray2013} that compare and catalog them. 
Among others, \citet{Bhattacharya2011} proposed a generalized form of the \citet{ShethTormen1999} function that we use here. Note that this generalization is not theoretically motivated by the excursion set formalism.

The multiplicity function from \citet[][equation 12-18]{Bhattacharya2011} is
\begin{equation}
\begin{aligned}
f_{Ba}(\sigma, z, \bar{A}, \bar{a}, \bar{p}, &\bar{q}) =  \bar{A}(z) \sqrt{\frac{2}{\pi}}\left[1+\left(\frac{\sigma^2}{\bar{a}(z) \delta^2_c}\right)^{\bar{p}(z)}\right] \cdots \\
& \cdots \left( \frac{\sqrt{\bar{a}(z)}\delta_c}{\sigma}\right)^{\bar{q}(z)} \exp\left[ -\frac{\bar{a}(z)}{2} \frac{\delta^2_c}{\sigma^2} \right].
\label{eqn:mf:model}
\end{aligned}
\end{equation}
In the case, $\bar{q}=1$, the parameters of Eq. (\ref{eqn:mf:model}) are the same as that of Eq. (\ref{eqn:mf:model:st02}) i.e. $\bar{A}=A$, $\bar{a}=a$, $\bar{p}=p$. The addition of the $\bar{q}$ parameter is strictly speaking not physically motivated, but provides a better fit to the data, see further down in the paper.

We then use the formalism of \citet{HU2003,Bhattacharya2011} to account for the large scale halo bias and the mass function's covariance.

\subsection{Large scale halo bias}
The large scale halo bias function is written in terms of the conditional, the unconditional mass function and a Taylor expansion \citep{ShethTormen1999,Bhattacharya2011}. This allows its formulation with the same parameters as the mass function 
\begin{equation}
\begin{aligned}
b(\sigma, z, \bar{a}, \bar{p}, \bar{q}) = & 1 + \frac{\bar{a}(z) (\delta^2_c/\sigma^2) -\bar{q}(z)}{\delta_c} \cdots \\
& \cdots + \frac{2\bar{p}(z) /\delta_c}{1 + (\bar{a}(z) (\delta^2_c/\sigma^2))^{\bar{p}(z)}}.
\label{eqn:bias:model}
\end{aligned}
\end{equation}
\subsection{Covariance matrix}
To model the covariance, we slightly adapt the notations from \citet{HU2003,Bhattacharya2011} as follows.

Let $\bar{\rho}$ be the average density of halos. 
We assume the over density of halos at a position $(z, \vec{x})$, denoted $\delta_{\textrm{halo}}(\sigma, z, \vec{x})$, to be related to the total mass density field $\delta_{DM}(\vec{x})$ by a biasing function, $b(\sigma, z)$. Note that, on large scales, this function is the bias mentioned in the previous Section.
\begin{equation}
\delta_{\textrm{halo}}(\sigma, z, \vec{x}) = b(\sigma, z) \delta_{DM}(\vec{x}).
\end{equation}
Then, within a window $W_a$, the average number density of halos, $n_a$ is given by
\begin{equation}
n_a(\sigma, z) = \bar{\rho} \int d\vec{x} \; W_a(\vec{x}) \;  b( \sigma_a, z_a) \delta_{DM}(\vec{x}).
\end{equation}

The covariance between the number densities $n_a(\sigma_a, z_a)$ and $n_b(\sigma_b, z_b)$ within in the windows $W_a$ and $W_b$ has two components: the shot noise variance, proportional to the inverse of the density times the volume $\sim(\bar{n} V)^{-1}$, and the sample variance: 
\begin{equation}
\begin{aligned}
 \frac{\langle n_a n_b \rangle - \bar{n}_a\bar{n}_b}{\bar{n}_a\bar{n}_b}& =  b(\sigma_a, z_a) D(z_a) b(\sigma_b, z_b) D(z_b)  \cdots  \\ & \cdots \times \int \frac{3 d^3k}{(2\pi)^3} W_a(k\, R_{box,\, a}) W^*_b(k\, R_{box,\, b}) P(k),
\label{eqn:variance:a}
\end{aligned}
\end{equation}
where $D$ is the growth factor, $V$ the volume of the box, $R_{box}=(3 V/4\pi)^{1/3}$ and $P(k)$ the dark matter power spectrum. We use a top-hat window functions.
The growth factor and the integral depend only on the cosmological model (and redshift) but not on the mass function model. 
The model of the bias function is directly related to the halo mass function model. Therefore once the mass function parameters are determined, the covariance matrix should be predictable.
Also, we note how the large-scale structure makes number counts of halos in distinct volumes covary. 
Our model of the covariance matrix is 
\begin{equation}
C_{\textrm{model}}(\sigma_a, \sigma_b) = \frac{Q}{\sqrt{\bar{n}_a \bar{n}_b } (V_a+V_b) } + \left( \frac{\langle n_a n_b \rangle -\bar{n}_a\bar{n}_b}{\bar{n}_a\bar{n}_b} \right),
\label{eqn:cv}
\end{equation}
where the $Q$ factor depends on the simulation size. 
This factor allows us to rescale small-sub-boxes estimates of the covariance to much larger computational simulations. We find the factor by observing how covariance scales with the box size. 
In the next section, we find that $Q=-3.62+4.89\log_{10}(L_{box} [h^{-1} Mpc])$ accounts well for all of the estimated covariance matrices, see Fig. \ref{fig:cov:fsnfsv}. 

\section{Simulations}
\label{sec:data}

The MultiDark simulation suite\footnote{\href{https://www.cosmosim.org/}{cosmosim.org}} is currently the largest public data base of high-resolution large volume boxes with $\sim4000^3$ particles. The simulations were run in the Planck cosmology \citep{Prada2012,Klypin2016} in a flat $\Lambda$CDM model with the $\Omega_m=0.307$, $\Omega_{\Lambda}=0.693$, $\Omega_b=0.048$, $n_s=0.96$, $h=0.6777$, $\sigma_8=0.8228$ \citet{Planck2014}. They provide, halos plus subhalos for all written outputs and for some boxes merger trees are also available.
We found three other relevant simulation sets to be compared with our study. 
\citet{Despali2016} is the current state-of-the-art halo mass function in Planck cosmology. 
They ran a suite of $1024^3$ particle simulations with different volumes and analyzed the mass function up to redshift 1.25. 
The DarkSkies simulations discussed in \citet{Skillman2014}, also run in Planck cosmology, used up to $10,240^3$ particles and cover much larger volume, though the current data release only provides data at redshift 0. The exact cosmological parameters differ a little from the ones used in MultiDark and \citet{Despali2016}. \citet{Ishiyama2015} provide a new suite of simulation in Planck cosmology, the largest simulation (of interest for this analysis) is not yet publicly available, so we did not include their data in the analysis. 
Other simulations covering large volumes with large amount of particles exist \citet[e.g.][]{Angulo2012,Heitmann2015}, but they were run in a different cosmology setup and are not yet publicly available. For completeness, also we mention the P-Millennium $\sim4000^3$ simulation although it is not publicly documented and released yet. 
In this study, we therefore use only the MultiDark simulations and the redshift 0 data produced by the DarkSkies simulation. These datasets constitute a non-negligible leap forward, for both resolution and volume, compared to the data used in \citet{Despali2016}. Table \ref{table:multidark:summary} summarizes and compares the main parameters of each simulation: length of the boxes, number of particles, force resolution, particle mass and number of snapshots. We note the latest advances in software enabling 20,000$^3$ particle simulations to converge in reasonable computing time \citep{PKDGRAV32016}.

We use a set of snapshots from each simulation to sparsely and regularly sample the redshift range $0<z<2.5$, \textit{i.e.} to cover the extent of galaxy surveys. Table \ref{table:multidarkonly} gives the number of snapshots used per simulation in our analysis. 

\begin{table*}
\centering
 \caption{Basic parameters of the simulations. $L_{box}$ is the side length of the simulation cube. $N_p$ is the number of particles in the simulation. $\epsilon$ is the force resolution at redshift $z=0$. $M_p$ is the mass of a particle. Ns is the number of snapshots available. The $\sigma_8$ column gives the input value and its measured deviation at redshift $z=0$. 
The column `cosmo' refers to the cosmology setup used to run the simulation: (a) refers to \citet{Planck2014} and (b) to \citet[WMAP,][]{WMAP72011}. 
The column `ref' gives the reference paper for each simulation: 
(1) stands for \citet{Klypin2016} $h$=0.6777, $\Omega_m=0.307$, 
(2) for \citet{Skillman2014} $h$=0.6846, $\Omega_m=0.299$, 
(3) for \citet{Despali2016} $h$=0.677, $\Omega_m=0.307$, 
(4) for \citet{Heitmann2015} $h$=0.71, $\Omega_m=0.27$, 
(5) for \citet{Angulo2012} $h$=0.73, $\Omega_m=0.25$. 
(6) for \citet{Springel2005} $h$=0.73, $\Omega_m=0.25$. 
(7) for \citet{Ishiyama2015} $h$=0.68, $\Omega_m=0.31$. 
A dash, `-', means information is the same as in the cell above. An empty space means the information is not available. 
The column nickname give the naming convention used throughout the paper, figures and captions.}
\label{table:multidark:summary} 
  \begin{tabular}{c rrrcc ccccc cc }
 \hline \hline
Box &  \multicolumn{4}{c}{setup parameters}&  Ns & $\sigma_8$ & cosmo & ref & nickname \\
 Name    &  $L_{box}$  & $N_p^{1/3}$ & $\epsilon$ & $M_p$ &  &  input, measured & \\
         &  \mpc &               &  \kpc &  \Msun &  \\
\hline 
SMD         & $590.2$  		& $3,840$ 	& 2.2 	& $1.4\times10^8$ 			& 88 & 0.8228, $-2.8\%$ 	& (a) & (1) & M04 \\
MDPL     	& $1,475.5$ 	& $3,840$ 	& 7.3 	& $2.2\times10^9$ 			& 128	& -, $+0.2\%$ &- &- & M10 \\
BigMD     	& $3,688.9$ 	& $3,840$	& 14.7 	& $3.5\times10^{10}$ 		& 80 	& -, $+0.5\%$	&- &-& M25 \\
BigMDNW 	& $3,688.9$ 	& $3,840$ 	& 14.7 	& $3.5\times10^{10}$ 		& 1 	& -, $+0.5\%$	& -& -& M25n  \\
HMD 		& $5,902.3$		& $4,096$ 	& 36.8 	& $1.4\times 10^{11}$  	& 128	& -, $+0.4\%$&-&- & M40 \\
HMDNW 		& $5,902.3$  	& $4,096$ 	& 36.8 	& $1.4\times 10^{11}$   	& 17 	& -, $+0.4\%$	& -& -& M40n \\
\hline
DarkSkies  	& $11,627.9$ 	& $10,240$ 	& 53.4 	& $5.6\times10^{10}$ 	&16& 0.8355, +0.0\% &(a)& (2) & D80 \\
-, 			& $2,325.5$ 	& $4,096$  	& 26.7 	& $7.1\times10^9$ 		&- &- &-  &- &DS\\
 -, 		& $1162.7$ 		& $4,096$ 	& 13.3 	& $8.8\times10^8$ 		&- &- & - &- &- \\
 -, 		& $290.7$ 		& $2,048$ 	& 6.7 	& $1.1\times10^8$ 		&- &- & - &- &-  \\
 -, 		& $145.3$ 		& -		 	& 3.3 	& $1.3\times10^7$ 		&- &- & - &- &-   \\
\hline
Ada 		& $92.3$ 		& $1,024$ 		&2.2  	& $2.8\times10^7$ 		&15& 0.829, & (a) & (3) & De\\
Bice 		& $184.6$ 		& - 			& 4.4	& $2.2\times10^8$		&15 & -,&- & -&-\\ 
Cloe 		& $369.2$ 		& - 			& 8.8	& $1.8\times10^9$ 		&15&  -,&- & -&-\\
Dora 		& $738.5$ 		& - 			& 17.7	& $1.4\times10^{10}$	&15 & -,&- & -&-\\
Emma 		& $1,477.1$ 	& - 			& 35.4	& $1.1\times10^{11}$ 	&15& -,&- & -&-\\
Flora 		& $2,954.2$ 	& - 			& 70.9	& $9.3\times10^{11}$ 	&15& -,&- & -&-\\
\hline
$\nu^2$GC-L	& $1647.0$ 	& $8,192$ 	&  		& $3.2\times10^8$ 		&  & 0.83 & (a) &  (7) & $\nu^2$GC \\
$\nu^2$GC-M	& $823.5$ 	& $4,096$ 	&  		& $3.2\times10^8$ 		& 4 & - & (a) &  (7) & - \\
$\nu^2$GC-S	& $411.7$ 	& $2,048$	&  		& $3.2\times10^8$ 		& 4 & - & (a) &  (7) & - \\
$\nu^2$GC-H1& $205.8$  	& $2,048$	&  		& $4.0\times10^7$ 		& 4 & - & (a) &  (7) & - \\
$\nu^2$GC-H3& $205.8$ 	& $4,096$	&  		& $5.0\times10^6$ 		& 2 & - & (a) &  (7) & - \\
$\nu^2$GC-H2& $102.9$ 	& $2,048$ 	&  		& $5.0\times10^6$ 		& 4 & - & (a) &  (7) & - \\
\hline
p-Millennium& $800.0$     	&   				&  		& $1.5\times10^8$ 		& 271 &  & (a) &  In prep. & P-Mi\\
\hline \hline
OuterRim 	& $4,225.3$ & $10,240$ 		& 7.0 	& $2.6\times10^9$ 			& 34& 0.84, & (b) &  (4) &OR \\
QContinuum	& $1,830.9$ & $8,192$ 		& 2.8 	& $2.1\times10^8$ 			&- 	& -  &-	&- &QC \\
Millennium XXL & $4,109.6$ & $6,720$ 	& 13.7 & $1.1\times10^{10}$ 		& & 0.9 & other & (5) & Mi-XXL\\
Millennium 	& 	$684.9$   	&	$2,160$ &  		& $1.1\times10^9$ 			& & - & - & (6) & Mi \\
\hline
\end{tabular}
\end{table*}

\begin{table}
\centering
 \caption{More parameters for the MultiDark simulation data used in this paper. The number of snapshots used in the analysis is the one that has a distinction between central and satellite halos, which is a subsample of the complete simulations.}
 \label{table:multidarkonly}
 \begin{tabular}{cc ccc }
 \hline \hline
Box &  \multicolumn{3}{c}{Number of snapshots with parent ids}\\
     &   all & $z<3.5$ & $z<2.5$ \\
\hline 
M04      	& 9 & 9 & 8 \\
M10     	& 11 & 11 & 10 \\
M25     	& 10 & 10 & 9 \\
M25n 	& 1 & 1 & 1 \\
M40 		& 128& 67 & 56  \\
M40n 	& 17 & 15 & 13 \\
\hline
\hline
\end{tabular}
\end{table}

The $RMS$ amplitude of linear mass fluctuations in spheres of 8 $h^{-1}Mpc$ comoving radius at redshift zero, denoted $\sigma_8$, holds a particular role when characterizing the abundance of halos. 
To have a more accurate estimate of the actual $\sigma_8$ in the simulation, we compare the dark matter power spectrum at redshift 0 measured in each simulation with the predicted linear power spectrum in the same cosmology. 
The mean of the square-root of this ratio evaluated on scales where the linear regime dominates gives the relative variation of the value of $\sigma_8$. 
We find variation smaller than $\sim$2\%; see Table \ref{table:multidark:summary}. 
In the following, we compute the mass -- $\sigma(M)$ relation using the measured value of $\sigma_8$ in each simulation. 
To compute these relations, we use the package \citet[][\textsc{HMFcalc}\footnote{\href{http://hmf.icrar.org/}{hmf.icrar.org}}]{Murray2013}.

\begin{figure*}
\begin{center}
\includegraphics[width=19cm]{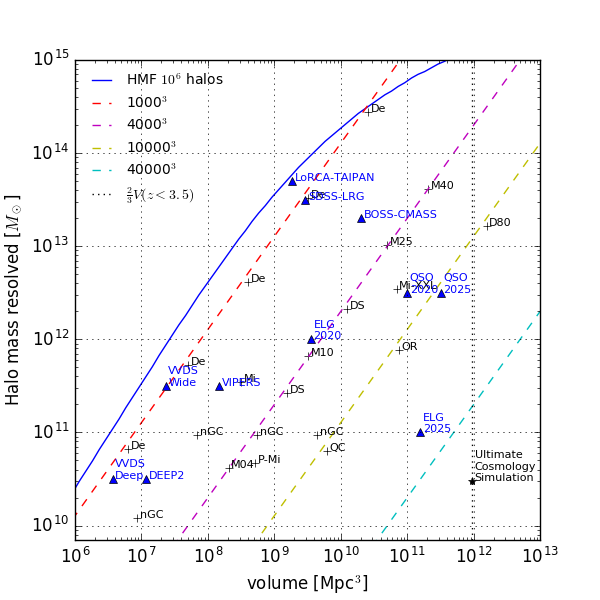}
\caption{Resolved halo mass vs. volume. The resolved halo mass is taken as 300 times the particle mass. 
The set of simulations discussed in this paper (black crosses, De: \citet{Despali2016}; M04, M10, M25, M40: MultiDark; DS, D80: DarkSkies; OR: OuterRim; QC: QContinuum; Mi: Millennium; nGC: $\nu^2$GC) are compared to current and future spectroscopic galaxy surveys (blue triangles). The galaxy surveys are tentatively placed according to halo mass values obtained with HOD models, the location is therefore not accurate but rather informative. 
Dashed diagonal lines relate the volume to the halo mass resolved assuming a constant number of particles $1000^3$ to $40,000^3$.
Assuming a halo abundance matching model, a simulation encompasses a galaxy sample located above and leftwards to its marker. 
We show a prediction of the redshift zero cumulative halo mass function (blue curve). 
It is the mass of the least massive halo among the 1,000,000 most massive halos expected in a simulation of the volume given in the x-axis.  
The total comoving volume of our past light cone within redshift 2.5 is $\sim10^{12}$ Mpc$^3$, the right boundary of the plot.}
\label{Fig:summaryBoxes}
\end{center}
\end{figure*}

To visualize the challenges of bridging the gap between $N$-body simulations and galaxy survey, we designed Fig. \ref{Fig:summaryBoxes}. 
In this figure, we compare existing simulations with observed galaxy surveys in the resolved halos mass vs. comoving volume plane. 
We consider the resolved halo mass to be 300 times the particle mass of a simulation. The total comoving volume of our past light-cone within redshift 3.5 projected on two third of the sky is $\sim10^{12}$ Mpc$^3$, the right boundary of the plot. 
We place the simulations enumerated in Table \ref{table:multidark:summary} according to their resolved halo mass and total volume (black crosses). 
We show with a set of dashed lines the relation between number of particles, volume and halo mass resolved. 

It shows how simulations progressed and our future needs (black star on the bottom right), from the top-left to the bottom-right. 
We show a prediction of the redshift zero cumulative halo mass function. 
It is the mass of the least massive halo among the 1,000,000 most massive halos expected in a simulation of the volume given in the x-axis. 
For example, in a volume of 10$^9$ Mpc$^3$, there are a million halos that have $M_{vir}>4\times 10^{13} M_\odot$. 
The galaxy surveys (blue triangles) are tentatively placed according to halo mass values obtained with HOD models. Given the uncertainty on the HOD model parameters, the halo mass value used could shift around by say a factor of 2 or 3. The survey volumes are accurate. The galaxy surveys represented are \citep[VIPERS,][]{Marulli2013}, \citep[VVDS-Wide,][]{Coupon2012}, \citep[VVDS-Deep,][]{Meneux2008}, \citep[DEEP2,][]{Mostek2013}, \citep[SDSS-LRG,][]{Padmanabhan2009}, \citep[BOSS-CMASS,][]{RodriguezTorres2015}, \citep[ELG 2020,][]{comparat_2013_bias,favole_2015_elg}, \citep[ELG 2025,][]{DESI2016}, \citep[QSO 2020,][]{QSO_sergio_2016}, \citep[QSO 2025,][]{DESI2016}. 
If a simulation point is to the lower right of a data point, it means the simulation is sufficient to construct at least one realization of the observations (assuming a halo abundance matching model). 
We note the challenge to simulate upcoming ELG samples to be observed by DESI, 4MOST, Euclid. Indeed a simulation with $L_{box}\sim 10,000 h^{-1}$ Mpc sampled with $\sim 20,000$ cube particles is needed. 
It seems that such simulations should become available in the coming decade. 
However, we do not need to simulate in a single box the exact volume of the observations to extract the cosmological information, see \citet{Klypin2017} for an extended discussion on the subject.

\subsection{Halo catalogs}
The halo finding process is a daunting task and in this analysis, we do not enter in this debate \citep[see][for a review]{Knebe2011,Knebe2013,Behroozi2015}. 
For the present analysis, we use the \textsc{rockstar} (Robust Over density Calculation using K-Space Topologically Adaptive Refinement) halo finder \citep{Behroozi2013}. 
Spherical dark matter halos and subhalos are identified using an adaptive hierarchical refinement of friends-of-friends groups in six phase-space dimensions and one time dimension. 

\textsc{rockstar} computes halo mass using the spherical over densities of a virial structure. 
Before calculating halo masses and circular velocities, the halo finder removes unbound particles from the final mass of the halo. 
We use halos that have a minimum of a $1,000$ bound particles, a very conservative threshold for convergence (some analysis use halos with $300$ particles, or even down to only 30 particles or so in the case of FoF halos). 
We characterize the halo population with two properties, $M_{vir}$ and $V_{max}$ at present.

For the halo mass, we use M$_{vir}$, defined relatively to the critical density $\rho_c$ by
\begin{equation}
\rm M_{vir}(z) = \frac{4\pi}{3} \Delta_{vir}(z) \Omega_m(z) \rho_c(z) R^3_{vir}.
\end{equation}
Indeed the halo $M_{vir}$ function was found to be closest to an eventual universal mass function \citep{Despali2016}. 
Throughout the analysis, we convert the mass variable to $\sigma$ as defined in Eq. (\ref{eqn:m:sigma})
To do so, we measure the dark matter power spectrum ($P_{DM}$) on each simulation at redshift 0. Then, we take the mean of the ratio $P_{DM} / P_{lin}$ on large scales; where $P_{lin}$ is the predicted linear power spectrum by CAMB using the cosmological parameters of the simulation. Finally, we rescale the M -- $\sigma$ relation accordingly to align all simulations to the input cosmological parameters. The value of the rescaling is given in the $\sigma_8$ column of Table \ref{table:multidark:summary}.

The maximum of the circular velocity profile is a measure of the depth of the dark matter halo potential well. It is expected to correlate well with the baryonic component of galaxies such as the luminosity or stellar mass as followed from the Tully-Fisher relation \citep{1977A&A....54..661T}. 
The maximum circular velocity is defined by Eq. (\ref{vmax:definition}). It has a very small dependence on radius and is therefore robustly determined, 
\begin{equation}
V_{max}=\max_r\left(\sqrt{\frac{GM(<r)}{r}} , \; \mathrm{over \; radius \; r} \right).
\label{vmax:definition}
\end{equation}

\subsection{Measurements}
\label{sec:mass:1pt}

We divide each snapshot in $1,000$ sub-volumes (on a grid of $10x10x10$). We compute the histogram of the halo mass in each sub-volume. 
The bins start at 8 and run to 16 by steps of $\Delta \log_{10}^M=0.05$. We denote, ${\rm N^{bin\, i}}$, the number count in a sub-volume in a mass bin. 
\citet{Lukic2007,Bhattacharya2011} corrected the mass assignment according to the force resolution of each simulation. We follow their corrections: $M_{corrected}=[1-0.04(\epsilon/650\,kpc)]\, M_{halo\; finder}$. The masses were overestimated by 0.3, 0.3, 0.1, 0.1, 0.05, 0.02 per cent in the M40, M40n, M25, M25n, M10, M04, respectively.

\begin{figure}
\begin{center}
\includegraphics[type=png,ext=.png,read=.png,width=7.cm]{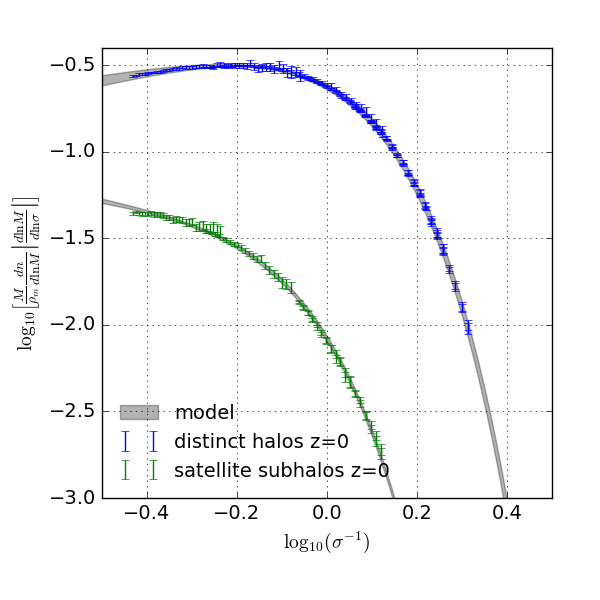}
\includegraphics[type=png,ext=.png,read=.png,width=7.cm]{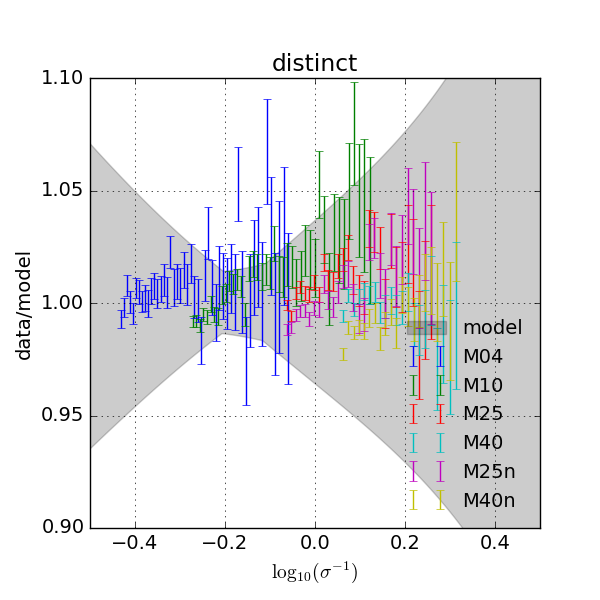}
\includegraphics[type=png,ext=.png,read=.png,width=7.cm]{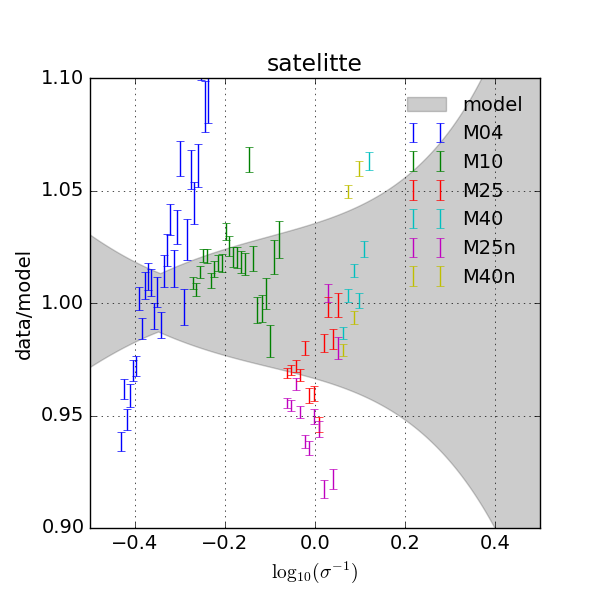}
\caption{Measurements of the differential halo $M_{vir}$ function for distinct halos as a function of $\log_{10}(\sigma^{-1})$ for the MultiDark simulations at redshift 0. The grey contours represents the best-fit models discussed in Section \ref{sec:resultMvirFunction}. The mean of the residuals for the distinct (satellite) halo mass function is 0.8\% (0.4\%) and the standard deviation of the residuals is 1.6\% (4.2\%) are shown in the middle (bottom) panel. It means the fit is very close to the data for the distinct halos and a little further for the subhalos.}
\label{Fig:mvir:fun:data}
\end{center}
\end{figure}

We estimate the uncertainty on the mass function using jackknife re-samplings by removing 10 per cent of the sub-volumes. We obtain 10 mass function estimates based on 90\% of each volume. 
In each simulation snapshot, we select bins where the halo mass is greater than a 1000 times the particle mass and where the number of halos is greater than 1000. We divide the number counts by the volume to obtain number densities
\begin{equation}
\rm dn(M) = \frac{{\rm N^{bin\, i}}( \log_{10}^{bin}(M_i) )}{{\rm Volume}},
\end{equation}
that we further divide by the natural logarithm of the bin width, to estimate the mass function, denoted interchangeably
\begin{equation}
n(\sigma,z)=\frac{dn}{d\ln M}.
\end{equation} 
The resulting mass function estimation for distinct and satellite halos at redshift 0 are presented in Fig. \ref{Fig:mvir:fun:data}. 
The measurements span the range ${\rm 11<\log_{10}(M_{vir}/M_\odot)<15(13.5)}$ for the distinct (satellites) halos. 

We find the DarkSkies halo mass function at redshift 0 to be 2\% lower than the combined MultiDark mass function. This is due to the lower matter content in the DarkSkies simulation. Also due to its large volume, the resolution does not enable to follow the mass function leftward of its knee, which prevents from fitting reliably the mass function models solely on the public DarkSkies data. The other DarkSkies simulations, that are smaller and complementary, are not provided to the public. Therefore, we do not push further the analysis with this simulation. 

\subsection{Covariance with mass}

\begin{figure*}
\begin{center}
\includegraphics[type=png,ext=.png,read=.png,width=7.cm]{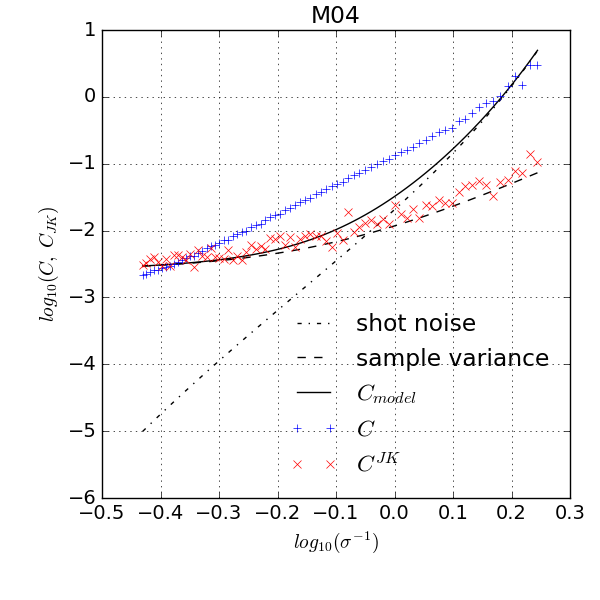}
\includegraphics[type=png,ext=.png,read=.png,width=7.cm]{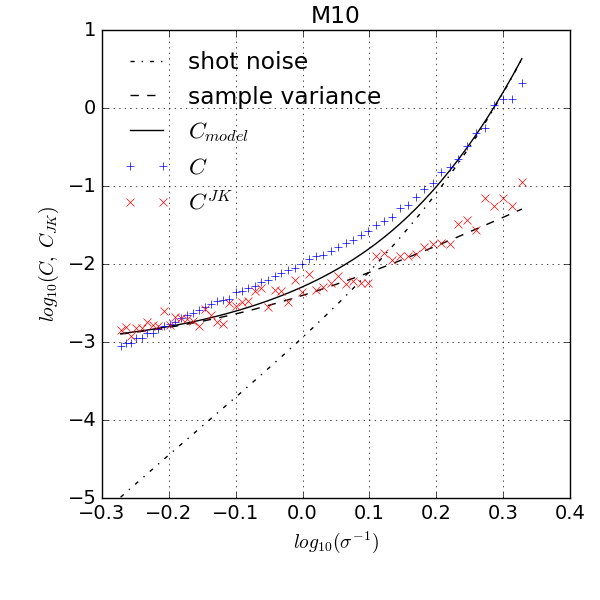}
\includegraphics[type=png,ext=.png,read=.png,width=7.cm]{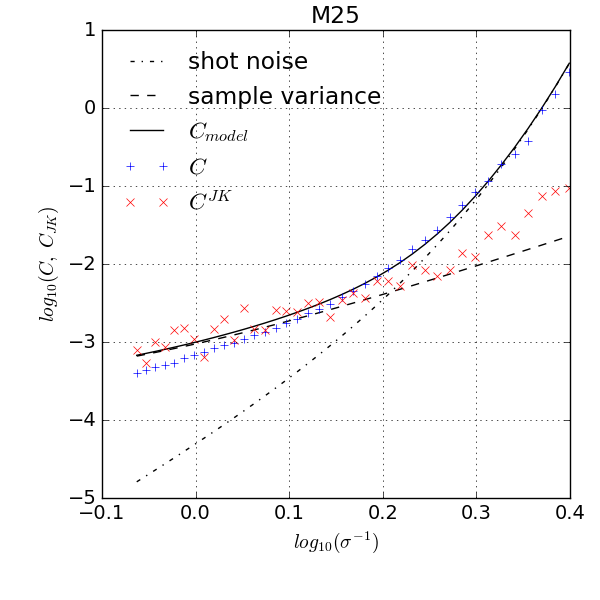}
\includegraphics[type=png,ext=.png,read=.png,width=7.cm]{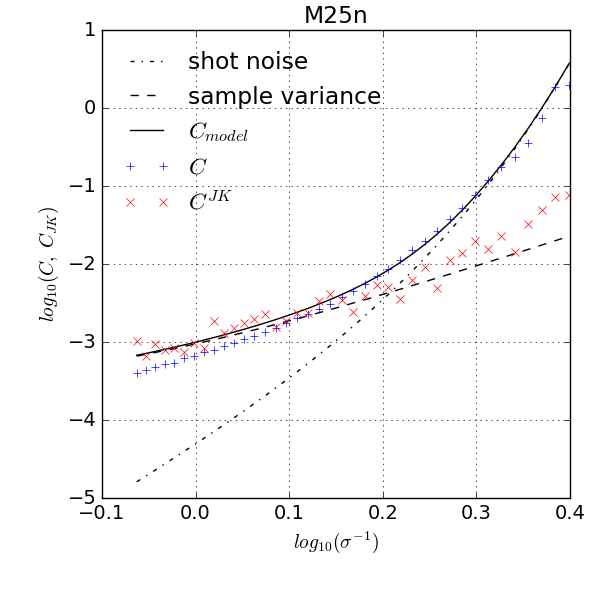}
\includegraphics[type=png,ext=.png,read=.png,width=7.cm]{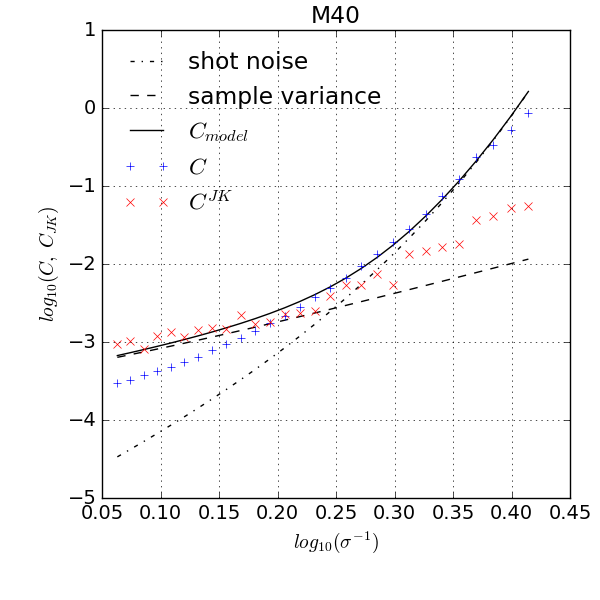}
\includegraphics[type=png,ext=.png,read=.png,width=7.cm]{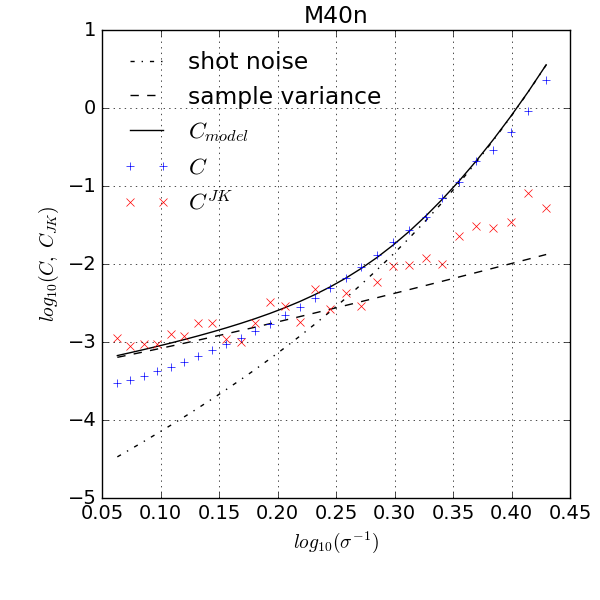}
\caption{Diagonal component of the covariance matrix measured in each MultiDark simulation (blue pluses) at redshift $z=0$ compared to the errors obtained via the jackknife method (red crosses). The model is decomposed into shot-noise and sample variance.}
\label{fig:covariance}
\end{center}
\end{figure*}

We construct two estimators of the uncertainty on the mass function measurements. We consider the redshift fixed. 
For both, we slice the simulations into $1,000$ sub-samples of equal volume. The grid is $10x10x10$. Each sub-sample has a volume $1,000$ times smaller than the initial simulation. 
The first method goes as follows. On each sub-sample, we estimate the mass function to obtain $N_{R}=1,000$ of them. We denote by $f_i(\sigma)$ the multiplicity functions deduced.
Then we compute the covariance matrix $C$ defined by
\begin{equation}
C(\sigma_a,\sigma_b)=\frac{\Sigma_i^{N_{R}}(f_i(\sigma_a) - \bar{f}(\sigma_a))(f_i(\sigma_b) - \bar{f}(\sigma_b))}{(N_{R}-1)},
\end{equation}
where $\bar{f}$ is the mean multiplicity function. Because each sub-sample ends up being quite small, the matrices hereby obtained do not cover a large dynamic range in mass. \newline
The second method is the jackknife. We group the sub-samples by batches of 100 to obtain $N_{R}=10$ realizations of the mass function using the complementary 900 sub-samples. 
The mass functions obtained are not independent, but they cover a larger mass range. From this method, we only infer the diagonal error
\begin{equation}
C^{JK}(\sigma)=\frac{\Sigma_i^{N_{R}=10}(f_i(\sigma) - \bar{f}(\sigma))^2}{(N_{R}-1)}.
\end{equation} \newline
We show the diagonal variances $C(\sigma,\sigma)$ and $C^{JK}(\sigma)$ on Fig. \ref{fig:covariance}. There is one panel per simulation snapshot at redshift 0. 
We note that both methods are in agreement when estimating the errors in the low mass regime. It is the regime where errors are dominated by sample variance. The jackknife method seems less sensitive to the shot-noise at the high-mass end. But this is simply a matter of the volume considered when estimating the uncertainty. Indeed in the jackknife method, we use 90\% of the volume whereas in the covariance, we only use 0.1\% of the volume. Therefore a factor of $\sqrt{1000}\sim30$ is expected between the two measurements. At the low-mass regime the sample variance seems underestimated by the full covariance method. This discrepancy cannot be explained by the difference in volume covered, we therefore assume this is a bias in the method.

The full covariance matrix varies smoothly with $\sigma$. The covariance matrix is not decreasing around its diagonal as the covariance matrix of the 2-point correlation function does \citep[see Fig. 7 of][]{comparat_2016_lorca}. Indeed there is a large amount of correlation between structure, i.e. the power spectrum of the dark matter is not zero. The model of the covariance matrix and its use in the analysis are discussed in Sec. \ref{sec:model}.

\subsection{Covariance with redshift}

\begin{figure*}
\begin{center}
\includegraphics[type=png,ext=.png,read=.png,width=8.5cm]{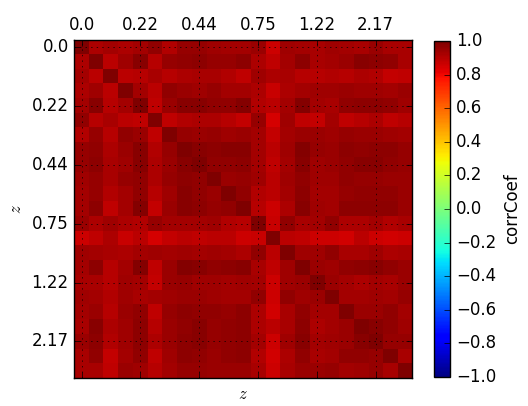}
\includegraphics[type=png,ext=.png,read=.png,width=8.5cm]{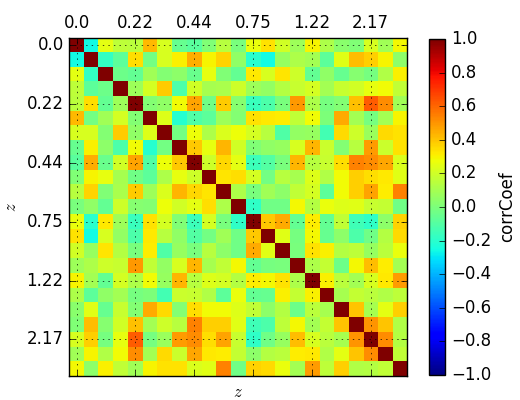}
\caption{$R_z(z_a,z_b)$, Redshift cross-correlation coefficient matrix of the number counts for density field values of $1+\delta=$ 100 (left) and 1000 (right).}
\label{fig:cov:zTrend}
\end{center}
\end{figure*}

\begin{table*}
\centering
 \caption{Parameters of the PPM-GLAM simulations run in Planck cosmology with $\sigma_8=0.8229$.}
\label{table:pm:summary}
 \begin{tabular}{c rrrcc ccccc cc }
 \hline \hline
Name    	&  $L_{box}$  & $N_p^{1/3}$ & $M_p$ 	& grid 	& dt  		& da & $N_R$\\
	        &  \mpc 			&  	             	&   \Msun &	& [Gyr] 	& \\
\hline 
pmA1        	& 737.7  	& 500 	&	$8.5\times10^{10}$	& 1,000 	& 0.5 & 0.0004 & 100\\
pmA2       	& 737.7  	& 500 	&	$8.5\times10^{10}$	& 1,000 	& 0.5 & 0.0002 & 100\\
pmA3        	& 147.5  	& 500 	&	$6.8\times10^{8}$	& 1,000 	& 0.5 & 0.0002 & 100\\
pmA4        	& 1475.5  	& 2,000 	&	$1.1\times10^{8}$	& 2,000 	& 0.5 & 0.0002 & 10 \\
\hline
pmB1        	& 1475.5  	& 1,000 	&	$8.5\times10^{10}$	& 1,000 	& 0.5 & 0.0002 & 10\\  
pmB2        	& 147.5  	& 1,000 	&	$8.5\times10^{7}$	& 1,000 	& 0.5 & 0.0002 & 10\\
pmB3       	& 14.7  		& 1,000 	&	$8.5\times10^{4}$	& 1,000 	& 0.5 & 0.0002 & 10\\
pmB4        	& 1.4  		& 1,000 	&	$8.5\times10^{1}$	& 1,000 	& 0.5 & 0.0002 & 10\\
\hline
\end{tabular}
\end{table*}

The mass function at redshift zero strongly depends on the mass function from previous redshifts i.e. on the complete formation history of the halos. 
Therefore fitting the redshift evolution of the parameters of the mass function is somewhat degenerate.
The additional information between two redshift bins are the new (sub)halos that formed, the mass increase of previous (sub)halos and the cross-talk between the two functions \citep[see][for an exhaustive list of events occurring during the evolution of the mass function]{Gi2010,vdbJ2016}. 
Due to the limited number of $N$-body realizations (6 for MultiDark), we cannot establish directly the redshift covariance of the mass function. \\
We run a set of approximate dark matter simulations to estimate the redshift covariance of the mass function to wisely choose the redshift sampling and avoid over-fitting in the later analysis. We run a set of Parallel Particle-Mesh GLAM simulations \citep[PPM-GLAM,][]{Klypin2017} with lower resolutions and lower time-step resolution than a typical high resolution $N$-body simulation to obtain a set of a 100 simulations with density field catalogs spanning the redshifts $0\leq z<3.2$ every 0.5 Gyr (23 time steps). With MultiDark, the number of realizations available is 6, a rather small number to obtain variances. On each realization and at each time step, we estimate the density field with a Cloud-In-Cell estimator. Table \ref{table:pm:summary} summarizes the PPM-GLAM runs. \newline
We estimate the redshift covariance matrix, $C_z$, of the density field function, $f^\delta$ as
\begin{equation}
C^\delta_z(z_a,z_b)=\frac{\Sigma_i^{N_{R}}(f^\delta_i(z_a)- \bar{f}^\delta(z_a))(f^\delta_i(z_b)- \bar{f}^\delta(z_b))}{(N_{R}-1)}
\end{equation}
at fixed values of the density field $\delta$. 
We deduce the Pearson product-moment correlation coefficients $R$ defined by
\begin{equation}
R_z(z_a,z_b)=\frac{C^\delta_z(z_a,z_b)}{\sqrt{C^\delta_z(z_a,z_a)C^\delta_z(z_b,z_b)}}.
\end{equation}\newline
The dark matter density field function, $f^\delta$, for $1+\delta = \rho/\bar{\rho}>10$ looks like a power-law. At the highest densities, $f^\delta$ is cut-off exponentially (due to finite resolution of the PPM-GLAM simulations). 
In the cross-correlation matrix, we find two regimes; see Fig. \ref{fig:cov:zTrend}. At the high-density field end, $1+\delta = \rho/\bar{\rho}>1000$, the cross-correlation coefficient is smaller than $<20\%$ between redshifts 0 and 10. The off-diagonal cross-correlations coefficient are of order of 10\%. Therefore each snapshot brings significant information in this regime. 
At the lower end of the density field function, $\delta = \rho/\bar{\rho}<200$,  the cross-correlation coefficient is larger than 80\%. It means that using a single redshift gives most of the information available. 
In between the transition is quite sharp, it suggests we should retain for the analysis the $z=0$ mass function measurements and the high-mass end of the $z>0$ mass function measurements. A cut-off at $\sim$200 times the density field seems reasonable. It corresponds to $\sim10^{12.9}M_\odot$. For simplicity, in this analysis we only use the redshift $z=0$ data and push back the question of accurate estimation of the redshift covariance for future studies.

These simulations give a sense of the redundancy of the information present in the data, but do not allow a robust estimation of the covariance matrix. With these simulations, we cannot weight each snapshot according to its information content. To do that, we would need a large amount of $N$-body simulations with halo finders run to estimate properly this covariance. Nevertheless it allows rejection of data with high covariance.\newline
Our understanding of the redshift covariance matrix is that the density field function at low over density is redundant with redshift.
We agree that between a density field function and a halo mass function there is a non-negligible step that is halo finding.
Nevertheless we think that adding all measured mass function points [in all written snapshots i.e. all the redshifts of the simulations] might lead to an incorrect statement as points cannot be considered to be strictly independent from one another.\newline
It seems that to further improve the accuracy of the halo mass function and in particular its evolution with redshift, we need to properly work out its redshift covariance matrix, but this needs significantly more simulations to be run, so we leave it for future studies.

\subsection{Large scale halo bias}

We compute the real space 2-point correlation function of the halo population in mass bins (identical as the ones used for the mass function) up separations to $r_{max}=20h^{-1}$Mpc. We follow a method described in \citet{martinezSaar2002} that goes as follows. \newline
We select all halos in a mass bin $[M, dM]$. It constitutes the complete sample of halos ($H_C$). 
Then, we select an 'inner' sample of halos ($H_I$) that are located at least $r_{max}$ away from any edge of the snapshot. 
We count all pairs between the $H_C$ and the $H_I$ sample using the \textit{scipy.spatial.ckdtree} python library \citep{scipyREF}. The histogram of the pair counts in bins of distance gives the number of pairs found at separation $r\pm dr/2$, denoted $N_\textrm{pairs}(r,dr)$. The real-space 2-point correlation function, $\xi$, is then obtained by
\begin{equation}
1 + \xi(r, dr, M, dM) = 
\frac{N_\textrm{pairs}(r,dr)}{\#H_C  \#H_I}
\frac{3 V_{snap}}{ 4\pi ((r+dr)^3 - r^3 )},
\end{equation}
where $V_{snap}$ is the volume of the snapshot and the distance binning parameter $dr=0.1$\Mpch. This is a fast and unbiased estimator of the 2-point function in simulations.\newline
We compute the redshift 0 linear correlation function, denoted $\xi^0_{lin}$, using CAMB and the Hankel transform \citep{Szapudi2005,Challinor:2011bk}\footnote{\href{https://pypi.python.org/pypi/hankel}{pypi.python.org/pypi/hankel}}.\newline
For scales $8<r<20$\Mpch, we divide the correlation function measured by the linear one. We take the mean to estimate the large scale halo bias 
\begin{equation}
b^2_h(M_{vir}) = \frac{1}{N_i} \sum_i \frac{\xi(M_{vir}, r_i)}{\xi^0_{lin}(r_i)}.
\end{equation}
We use the standard deviation of the latter ratio to estimate its uncertainty.

\begin{figure}
\begin{center}
\includegraphics[type=png,ext=.png,read=.png,width=7.8cm]{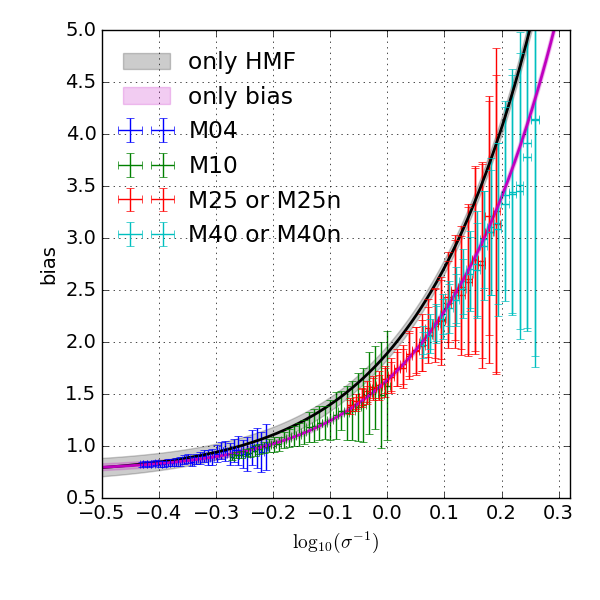}
\caption{Large scale halo bias vs. halo mass. Error bars show the data from the MultiDark simulations at redshift 0. The bias predicted using the best-fit parameters obtained on the HMF is shown in grey and the bias model fitted on the bias data is shown in magenta.}
\label{fig:halo:bias}
\end{center}
\end{figure}

Fig. \ref{fig:halo:bias} shows the halo bias measured at redshift 0 and the best fit models. The agreement between the data and the model is very good; see the discussion in the next Section.
  
\section{Results}
\label{sec:resultMvirFunction}
\begin{table*}
\centering
 \caption{Best-fit parameters of the model at redshift zero. $D(S)MF$ stands for distinct (satellite) mass function. B11: \citet{Bhattacharya2011}. D16: \citet{Despali2016}. A dash `-' means the entry is the same as above.}
\label{table:model:params}
\begin{tabular}{cccc ccccc }
\hline \hline
A(0) & a(0) & p(0) &  & $\chi^2/n\, d.o.f$ & $P(X>x, dof)$ & data & model Eq. & ref  \\ 
\hline
0.333$\pm$0.001 & 0.794$\pm$0.005 & 0.247$\pm$0.009 & & & & &  (\ref{eqn:mf:model:st02})& D16\\ 
0.3170$\pm$0.0008 & 0.818$\pm$0.003 & 0.118$\pm$0.006 &  & 238.69/ 187 = 1.28 & 0.7\% & MD DMF &-& this paper\\ 
0.0423$\pm$0.0003 & 1.702$\pm$0.010 & 0.83$\pm$0.04 &  &  31.03 / 84 = 0.37 & 100\% & MD SMF  & -& -\\ 
\hline\hline
$\bar{A}(0)$ & $\bar{a}(0)$ & $\bar{p}(0)$ & $\bar{q}(0)$ & $\chi^2/n\, d.o.f$ & $P(X>x, dof)$ & data & model Eq. & ref  \\ 
\hline
0.333 & 0.786 & 0.807 & 1.795 & & & & (\ref{eqn:mf:model})& B11 \\ 
0.280$\pm$0.002 & 0.903$\pm$0.007 & 0.640$\pm$0.026 & 1.695$\pm$0.038 & 138.76 / 186 = 0.75 & 99.6\%& MD DMF &-& this paper \\
0.27$\pm$0.02 & 0.92$\pm$0.03 & 0.36$\pm$0.68 & 1.6$\pm$0.6 & 9.13 / 21 =0.43 & 98.9\%& DS DMF &-& this paper \\
free  & 0.740$\pm$0.008 & 0.61$\pm$0.02 & 1.64$\pm$0.03 & 8.36/141 = 0.059 & 100\% & halo bias &-& this paper \\
\hline
\end{tabular}
\end{table*}

The determination of the best-fit model requires the assignment of errors on the data points. 
The covariance matrix discussed in the previous section is proportional to the product of the biases
\begin{equation}
C(\sigma_1,\sigma_2)\propto \frac{b(\sigma_1) b(\sigma_2)}{\sqrt{\bar{n}(\sigma_1)\bar{n}(\sigma_1)}}.
\end{equation}
Thus, each line of the matrix is proportional to another lines of the matrix, making it singular. It prevents from estimating the $\chi^2$ statistics for a given data-model pair, ($D$, $M$) via the inverse of the covariance matrix $\chi^2 = (D-M) \cdot C^{-1} \cdot (D-M)^T$.
 
We circumvent this issue as follows. 
First, in Sect. \ref{subsec:res:hmf} we use the uncertainty estimated with the jackknife method on the mass function and fit only the mass function data. Then in Sect. \ref{subsec:res:bias}, we fit the bias equation that involves the same parameters as the mass function to obtain another constraint on the parameters based on the covariance of the data. Finally in Sect. \ref{subsec:res:covariance} , we provide a relation to predict the covariance matrix for a given simulation.

\subsection{Distinct halo mass function}
\label{subsec:res:hmf}

To determine the best parameters for the mass function of distinct halos, we use a $\chi^2$ minimization algorithm\footnote{scipy.optimize.minimize: \href{https://docs.scipy.org/doc/scipy-0.18.1/reference/generated/scipy.optimize.minimize.html}{docs.scipy.org}} to obtain the set of best-fit parameters.
We fit the mass function model from Eqs. (\ref{eqn:mf:model:st02}) and  (\ref{eqn:mf:model}), to the data at redshift zero. We thus constrain the two sets of parameters  ($A, a, p$) and ($\bar{A}, \bar{a}, \bar{p}, \bar{q}$). 
We determined the parameters for different flavors of the data. `MD D(S)MF' stand for the distinct (satellite) mass function from MultiDark data. `DS DMF´ stand for the distinct mass function from DarkSkies data.  We use the Jackknife diagonal errors.
The fit of equation (\ref{eqn:mf:model:st02}) on the MD DMF gives a reduced $\chi^2=1.28$. The model is not a satisfying statistical representation of the data as the probability of acceptance is 0.7\%. We find parameters somewhat discrepant to what was found in \citet{Despali2016}. 
The fit of equation (\ref{eqn:mf:model}) to the MD DMF gives a reduced $\chi^2\sim0.75$, meaning it is an accurate description of the data. The probability of acceptance is $>99\%$. 
We find ($\bar{A}(0)$, $\bar{a}(0)$, $\bar{p}(0)$, $\bar{q}(0)$)=(0.280$\pm$0.002, 0.903$\pm$0.007, 0.640$\pm$0.026, 1.695$\pm$0.038). 
Table \ref{table:model:params} hands out the best-fit parameters obtained. 
We therefore think that adding the $\bar{q}$ parameter suggested by \citet{Bhattacharya2011} enhances significantly the quality of the fit to the $DMF$. The bottom panel of Fig. \ref{Fig:mvir:fun:data} shows the residuals after the fit of the model given in equation (\ref{eqn:mf:model}). 
The mean of the residuals for the distinct halo mass function is 0.8\% and the standard deviation of the residuals is 1.6\%. It means the fit on average underestimates the HMF by less than 1\%. Furthermore, except for a few outliers the MD DMF is very well described by the model to the $<2\%$ level. 

\begin{figure}
\begin{center}
\includegraphics[type=png,ext=.png,read=.png,width=8cm]{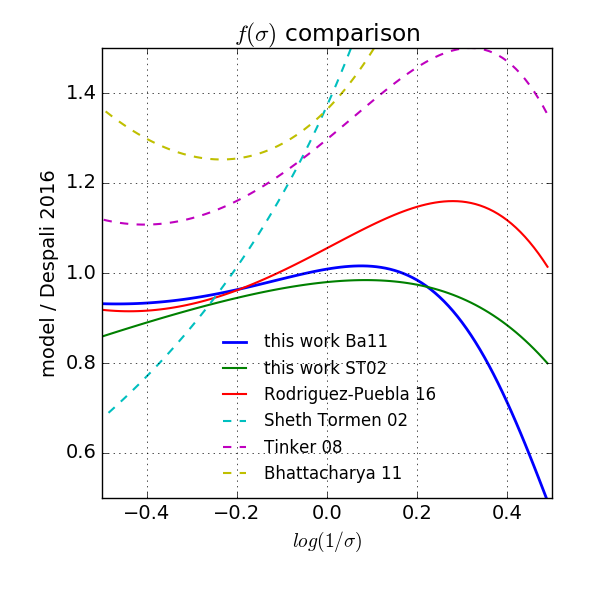}
\caption{Comparison of mass functions with respect to the \citet{Despali2016} fit. The line 'this work Ba11' corresponds to the fits of equation (\ref{eqn:mf:model}) to the data and 'this work ST02' corresponds to the fits of equation (\ref{eqn:mf:model:st02}) to the data. Studies done in the Planck cosmology have solid lines whereas studies in other cosmologies are shown with dashes. The difference at large masses is due to the difference in the simulation volumes.}
\label{fig:comparison:previous:fits}
\end{center}
\end{figure}

We compare our fits to previous ones in Fig. \ref{fig:comparison:previous:fits}. The mass function differs from up to a factor of two when compared to different cosmologies. Our fit agrees within $<$10\% with other analysis in a Planck cosmology in the lower mass regime. At larger masses, the disagreement between our measurements and previous ones in Planck cosmology is due to the difference in the data used. 
In this paper, we use extremely large simulations whereas in previous analysis, the largest simulation were covering volumes 8 to 64 times smaller. The high-mass end being modeled by an exponential, it drives the fit to a different location in parameter space.

\subsection{Large scale halo bias}
\label{subsec:res:bias}

The fit of the model given in Eq. (\ref{eqn:bias:model}) suggests the following set of parameters $(\bar{a}, \bar{p}, \bar{q})=(0.740\pm0.008, 0.61\pm0.02, 1.64\pm0.03)$. These are in slight tension with that of the halo mass function model (1 $\sigma contours do overlap$); see Table \ref{table:model:params} for a face-to-face comparison of the figures.
It is slightly higher for large masses and slightly lower for low mass. 

We are pleased to see that the excursion-set formalism works well to describe the mass function and the large scale halo bias precisely. Such a low level of tension is worth the praise. 

A joint fit to solve this issue is not straightforward. Indeed the large scale halo bias is related to the uncertainty on the mass function. We leave this for future studies. 

\subsection{Covariance matrix}
\label{subsec:res:covariance}

In the comparison of the diagonal errors estimated, see Fig. \ref{fig:covariance}, the two methods showed some disagreement: at the high-mass end where errors are dominated by the shot-noise and at the low-mass end where the errors are dominated by the sample variance. 
The difference in shot-noise is understood as the volumes used differ in the two error-estimating methods.
On the contrary, the difference in sample variance is puzzling. Indeed when using a larger volume, the sample variance estimated is higher than in the method using a smaller volume. This seems rather strange, as we expected the opposite. 
We take a conservative option. 
We consider the maximum of the two error estimates to fit the model: the $JK$ estimates at the low-mass end and the covariance at the high-mass end.

According to the model, fitting all the coefficients of the covariance matrix is redundant. 
The shot-noise component is a scaling relative to the inverse of the density times the volume. 
The sample variance depends on the product of the biases and on the cosmology. 
Therefore as soon as a single lines of coefficient of the covariance matrix is reproduced by the model, other coefficients should be in line with the model. 
This is indeed what we observe. 
As data points, we simply use the diagonal of the covariance matrix. Note that the points are for $L_{box} [h^{-1} Mpc] = $ 40, 100, 250 and 400, a factor of 10 smaller than the boxes used for the mass function estimate.

\begin{figure}
\begin{center}
\includegraphics[type=png,ext=.png,read=.png,width=8cm]{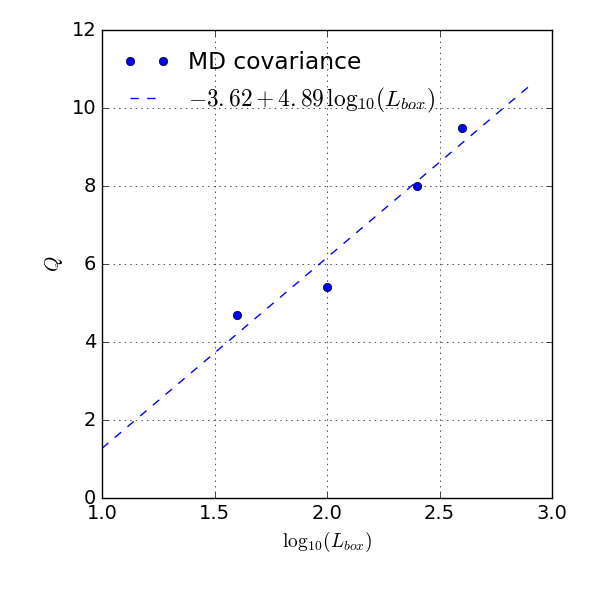}
\caption{$Q$ covariance rescaling factor vs. side length of the simulation and its linear fitting relation, see Eq. (\ref{eqn:fsn}).}
\label{fig:cov:fsnfsv}
\end{center}
\end{figure}

We fit a linear relation between the $Q$ factor and the log of the side length of the simulations (i.e. the length of the simulations divided by 10 due to the sub-sampling). The uncertainty on the coefficients of the covariance matrix is unknown, so we perform a fit where the data points are equally weighted. 
Using the large scale halo bias model from the previous subsection, we find that the following fitting relation,
\begin{equation}
Q=-3.62+4.89\log_{10}(L_{box} [h^{-1} Mpc]),
\label{eqn:fsn}
\end{equation}
produces a covariance matrix model very close to the MultiDark data at redshift 0. 
Figure \ref{fig:cov:fsnfsv} shows the $Q$ vs. the size of the simulation. 
We find the model to account well for the measured covariance, see Fig. \ref{fig:covariance} where the solid, dashed and dotted lines represent each component of the model. By combining equations (\ref{eqn:fsn}), (\ref{eqn:cv}) and $\sqrt{C_{\textrm{model}}(\sigma, \sigma, L_{box})}$, one predicts a reliable uncertainty on the distinct halo mass function for any simulation in the Planck cosmology.

\subsection{Subhalo and substructure mass function}

\begin{table*}
\centering
\caption{Number of distinct halos - subhalo pairs at redshift 0 split in distinct halo mass bins. Best-fit parameters for Eq. (\ref{eqn:progMF}) for each host halo mass bin are given below. The last column is the fit using all the data together.}
\label{table:subhalos}
 \begin{tabular}{c ccc ccc }
 \hline \hline

box & 12.5 - 13 & 13 - 13.5 & 13.5 - 14 & 14 - 14.5 & 14.5 - 15.5 	&	12.5 - 15.5\\
\hline 
M04   	& $515,922$ & $504,923$ & $441,228$ & $284,992$ & $144,352$ & 	$1,891,417$		\\
M10    	& $938,628$ & $879,394$ & $729,358$ & $480,041$ & $200,699$ & 	$3,228,120$		\\
M25    	& $788,780$ & $1,426,470$& $1,337,316$ & $833,535$ & $325,951$& $4,712,052$		\\
M25n 	& $784,519$ & $1,414,136$& $1,318,048$& $822,464$& $329,225$ & 	$4,668,392$		\\
M40 	& $19,793$  & $7,963,12$ & $1,619,226$& $1,199,845$& $467,090$ & $4,102,266$	\\
M40n	& $20,988$  & $797,074$ & $1,578,780$& $1,167,971$& $466,143$ &  $4,030,956$	\\
total   & $3,068,630$  & $5,818,309$  & $7,023,956$  & $4,788,848$  & $1,933,460$  & $22,633,203$ \\
\hline 
parameter & \multicolumn{6}{c}{best-fit values} \\
\hline
$-\alpha_{sub}$			& $1.73\pm0.03$ & $1.76\pm0.02$ & $1.78\pm0.01$ & $1.799\pm0.006$ & $1.834\pm0.004$ &  $1.804\pm0.004$\\
$\beta_{sub}$			& $5.34\pm0.16$ & $5.95\pm0.18$ & $6.12\pm0.16$ & $6.32\pm0.21$ & $5.87\pm0.27$ &  $5.81\pm0.09$\\
$-\log_{10}{N_{sub}}$ 	& $2.19\pm0.05$ & $2.15\pm0.03$ & $2.15\pm0.02$ & $2.25\pm0.01$ & $2.33\pm0.01$ &  $2.250\pm0.008$ \\
$\gamma_{sub}$			& $1.95\pm0.14$	& $2.28\pm0.11$	& $2.46\pm0.09$	& $2.62\pm0.09$	& $2.92\pm0.11$	& $2.54\pm0.05$\\
\hline
\end{tabular}
\end{table*}

In this analysis, we do not enter into the debate of the definition of subhalos. 
We use the subhalos as obtained by the \textsc{rockstar} halo finder at redshift zero. 
The substructure hierarchy in dark matter halos was investigated in details by \citet{Gi2010,vdbJ2016}. 
They argue two function are needed to fully characterize in a statistical sense the subhalo population: the halo mass function and the substructure mass function. The convolution of the two gives the subhalo mass function. 

We measure the subhalo mass function with the same method as for the distinct halo mass function; see Fig. \ref{Fig:mvir:fun:data}. 
We fit the subhalo mass function (MD SMF in Table \ref{table:model:params}) with equation (\ref{eqn:mf:model:st02}) and obtain a reduced $\chi^2\sim0.37$, meaning it is an accurate description of the data (Probability of acceptance 100\%). 
We find ($A(0)$, $a(0)$, $p(0)$)=(0.0423$\pm$0.0003, 1.702$\pm$0.010, 0.83$\pm$0.04). 
Adding an additional parameter $q$ is not necessary. 
The mean of the residuals for the subhalo mass function compared to this model is 0.4\% and the standard deviation of the residuals is 4.2\%. So the model is a little further away on average than for the MD DMF. To further refine the model, a complete discussion on what a subhalo is would be necessary. For the purpose of halo occupation distribution, adding a subhalo mass function with a 4\% precision is a non-negligible advance. We warn the reader that the excursion set formalism does not predict the sub clumps within halos. We simply use the function (\ref{eqn:mf:model:st02}) as an analytical model to describe the data.

\begin{figure*}
\begin{center}
\includegraphics[type=png,ext=.png,read=.png,width=7.5cm]{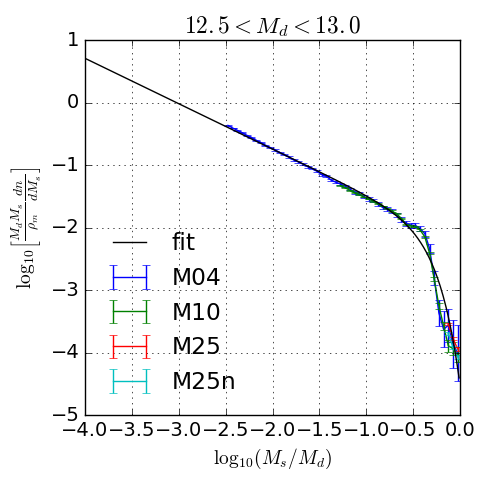}
\includegraphics[type=png,ext=.png,read=.png,width=7.5cm]{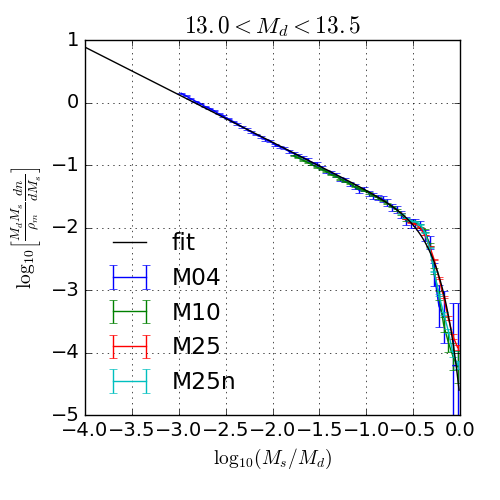}
\includegraphics[type=png,ext=.png,read=.png,width=7.5cm]{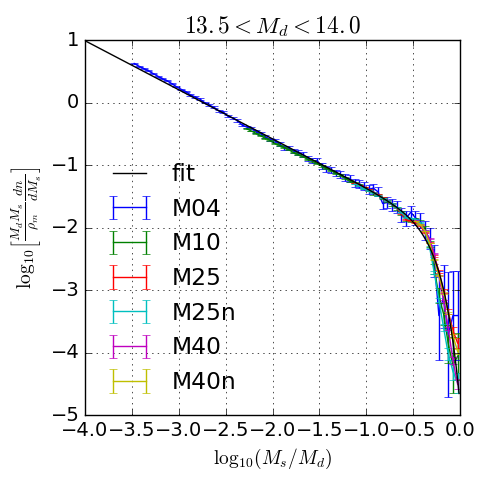}
\includegraphics[type=png,ext=.png,read=.png,width=7.5cm]{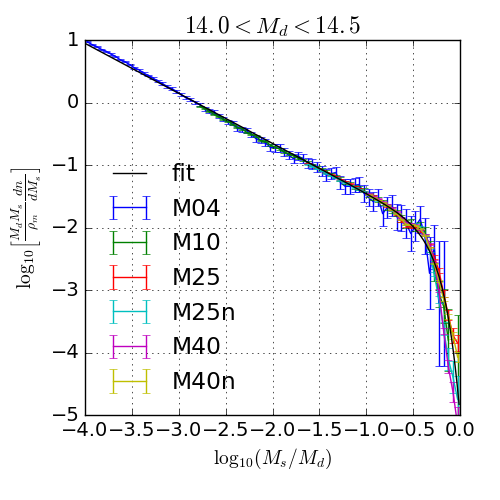}
\includegraphics[type=png,ext=.png,read=.png,width=7.5cm]{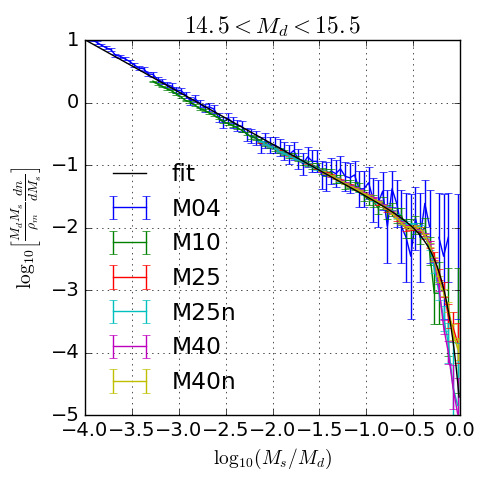}
\caption{Substructure mass function for five distinct (host) halo mass bins. The model seems quite independent of the host halo mass bin.}
\label{fig:subhalo:MFproj}
\end{center}
\end{figure*}

Then, for a subhalo of mass $M_s$ we consider its relation to its host, a distinct halo of mass $M_d$, by studying the distribution of the ratio $Y=M_s/M_d$. In this aim, we measure the so-called substructure mass function, defined by the left part of Eq. (\ref{eqn:progMF}) and shown on Fig. \ref{fig:subhalo:MFproj}.
\begin{equation}
\log_{10}\left[ \frac{M^2_d}{\rho_m} \frac{dn}{dM_s} \right] (Y)= N_{sub} Y^{\alpha_{sub}}  e^{-\beta_{sub} Y^{\gamma_{sub}} }.
\label{eqn:progMF}
\end{equation}
Note that $M_s$ is not the mass at the moment of accretion of the subhalo but the mass measured at redshift 0. We parametrize it similarly to \citet{vdbJ2016} with 4 parameters: overall normalization, $N_{sub}$, power-law at low mass ratio, $\alpha_{sub}$, and two parameters for the exponential drop: $\beta_{sub}$ and $\gamma_{sub}$.

The substructure mass function represents the abundance of subhalos as a function of the mass ratio between the subhalo and its host distinct halo (in a distinct halo mass bin); \citep[see equation (2) and Fig. 3 of ][]{Gi2010} and \citep[][equation (6) and Fig. 3]{vdbJ2016}. In these works, the authors consider a complete world model of how subhalos evolve. 
In this analysis, we focus on the practical aspect of a relation that given a halo population, one can predict the characteristics of its subhalo population. 
Therefore, we do not apply the exact same formalism as in previous works, but rather something more practical, at fixed redshift. 
We use the mean density of the Universe to obtain a dimensionless measurement, therefore the normalization parameters have a different meaning than in previous studies. 
Subsequently, we adjust a four-parameter model, given in the right part of Eq. (\ref{eqn:progMF}) to 5 host halo mass bins and to all the data simultaneously. 
Fig. \ref{fig:subhalo:MFproj} shows the substructure mass function measured at redshift 0 in the mass bins delimited by 12.5; 13; 13.5; 14; 14.5; 15.5. 
The parameters obtained are given in Table \ref{table:subhalos}. 
The $22,633,203$ subhalos-halo pairs considered constitute a sample that is more than an order of magnitude larger than any previous study. 
The power-law found is compatible with $-\alpha_{sub}=$-1.804$\pm$0.004 in every host mass bin. It confirms measurements from previous analysis, though with greater accuracy. The other parameters found are compatible between mass bins. To a good approximation, the parameters $-\alpha_{sub}=-1.8$, $\beta_{sub}=5.8$ and $-\log_{10}{N_{sub}}=2.25$, $\gamma_{sub}=2.54$ provide a good description of the substructure mass function (whatever the host halo mass bin).

\section{Summary and discussion}

In this analysis, we measured at redshift zero the mass function for distinct and satellite subhalos and the substructure mass function to unprecedented accuracy thanks to the MultiDark Planck simulation suite. Indeed these simulations encompass 8 times larger volumes than what was used in previous studies. We measured and modeled the large scale halo bias of the distinct halos. Then, we estimated for the first time the full covariance matrix of the distinct halo mass function with respect to mass. To refine our knowledge of the satellite subhalo population, we also estimated and modeled the substructure function.\newline
We find that the \citet{Bhattacharya2011} model is a good description for the measurements related to the distinct halo population: its mass function, its large scale bias and the covariance of the mass function. 
This new set of models for the mass function and for the velocity function should allow analytical halo occupation distribution models to reach better accuracy. We give practical fitting formula and their evolution with redshift of the $V_{max}$ function in Appendix.

\subsection*{Halo finding process}
The halo finding is a difficult task, reason being that, both the theoretical and the empirical definition of what a halo is, are not precise. \newline 
About the empirical definition of a halo. \citet{Knebe2011,Knebe2013,Behroozi2015} showed that when varying the halo finder on a single simulation, one should expect variations in the distinct halo mass function of the order of 10-20\%. 
This estimate, done on a rather small simulation (500\Mpch) with a small number of particles ($1024^3$), should be regarded today as an upper limit. 
Hopefully such an exercise will be repeated with current and future simulations to reach a better empirical halo definition. \newline 
About the theoretical halo definition, it seems recent investigations on the extended spherical collapse models by \citet{Popolo2017} point towards a modification of the \citet{ShethTormen1999} along the lines of the modifications made by \citet{Bhattacharya2011}. So there might be a physical reason behind the fact that the \citet{Bhattacharya2011} is a better description of the data than \citet{ShethTormen1999}.\\

Unlike distinct halos, the satellite subhalo definition has not yet reached a consensus in the community. Theoretical advances are pushing towards a unified subhalo model so this uncertainty should hopefully vanish soon \citep{vdbJ2016}. Nevertheless, we provided accurate fits of the statistics obtained with MultiDark combined with \textsc{rockstar}. \\ 

\subsection*{Redshift evolution of the mass function}
The redshift covariance of the density field function indicates that the debate about the universality of the mass function throughout redshift might be an ill-posed question.\\
Given the covariance between different redshift bins in the low-mass end of the density field function, it is hard to define properly how its evolution with redshift should be modeled. Simply using all the redshift outputs produced by the simulation is redundant. 
We therefore think the question of the universality needs be approached with a slightly different theoretical background. Many more N-body simulations would need to be run to obtain deep insights on the redshift covariance of the halo mass function. 
But it does not seems reasonable to run a thousand MultiDark of DarkSkies simulations ? 
To save computation time, a possibility would be to study the evolution of the density field with the new PPM-GLAM method. 
In this paradigm, the number of realizations is not an issue and cosmological parameters are easily varied. \\

\subsection*{About the effects of baryons on the halo mass function}
The baryons hosted by dark matter halos influence the total mass enclosed in the halo. 
Supernovae and active galactic nuclei feedbacks expel gas from the halo to the inter galactic medium. 
The total mass enclosed in halos where baryonic physics is accounted for is of order of 20\% or lower. 
Therefore the halo mass function estimated on dark matter only simulations suffers a bias. 
It seems the number density of DM only halos is greater than that of DM+baryon halos by a factor $\sim20\%$ at $M\sim10^{9}h^{-1}M_\odot$. In clusters the halo number densities seem in agreement. 
We summarize numbers obtained from various studies in Table \ref{table:baryon:effectMF}. 
At redshift 0, it seems there is a consensus for clusters (impact negligible) and halos with $log_{10}M<12$ (-20\% effect). The evolution of this effect with redshift is not clear. \citet{Vogelsberger2014MNRAS.444.1518V} and \citet{Schaller2015MNRAS.451.1247S} show an effect more or less constant with redshift. The most recent simulations \citep{Bocquet2016MNRAS.456.2361B} advocate the effect is negligible at redshift 2 and starts around redshift 1. Recently, \citet{Despali2016arXiv160806938D} tested these models by comparing with observed strong lensing events. 
With current statistics it does not allow to choose between feedback models, but with larger samples, the strong lensing probe should decide this problem.
Note that, the trend with mass vary from a simulation to another due to the differences in the AGN feedback or the supernovae model used. This result is indeed dependent on the recipe of AGN and supernovae feedback, so the true value could be larger (or smaller) but it is difficult to quantify by what amount.
 
\begin{table*}
\centering
 \caption{Ratio between the halo mass function with and without baryonic effect as a function of halo mass, $f_{hydro}/f_{DM\; only}$. References are 1:  \citet{Velliscig2014MNRAS.442.2641V}; 2: \citet{Vogelsberger2014MNRAS.444.1518V}; 3: \citet{Schaller2015MNRAS.451.1247S}; 4: \citet{Tenneti2015MNRAS.453..469T}; 5: \citet{Bocquet2016MNRAS.456.2361B}}
 \label{table:baryon:effectMF}
 \begin{tabular}{cccccccccccc }
\hline
\hline
 9-10 & 10-11 & 11-12 & 12-13 & 13-14 & 14-15 & simulation & reference \\
\hline
  &  & 0.8 & 0.8 & 0.8 & 0.9 & OWLS & 1 \\
 0.8 & 0.8 & 1.1 & 1 & 0.9 & 0.9 & Illustris  & 2 \\
 0.7 & 0.8 & 0.85 & 0.9 & 0.95 & 1 & Eagle & 3 \\
 0.8 & 0.85 & 0.85 & 0.9 & 0.95 & 1 & massive black 2 & 4 \\
 0.9 & 0.9 & 0.9 & 0.9 & 0.9 & 1 & Magneticum & 5 \\
\hline
\end{tabular}
\end{table*}

\subsection*{Outlook}

All in all it seems assuming a few percent statistical errors and of order of tens of percents systematical errors reasonably represents our current knowledge of the distinct halo mass function. To enable percent precision with mass function cosmology, these results call for deeper investigations. First about the redshift and mass covariances of the distinct halo mass function to be able to do proper statistical fits on the data. Second about seeking a better empirical and theoretical definition of what a dark matter halo is. Last about the remaining n-point functions that carry the next order of information about what halos are and how they behave.

\bibliographystyle{mnras}
\bibliography{biblio}

\section*{Acknowledgements}
\vspace{0.2cm}
JC thanks J. Vega, Sergio A. Rodr\'iguez-Torres, D. Stoppacher, A. Knebe, the eRosita cluster working group and the referee for insightful discussion or comments on the draft.
JC and FP acknowledge support from the Spanish MICINNs Consolider-Ingenio 2010 Programme under grant MultiDark CSD2009-00064, MINECO Centro de Excelencia Severo Ochoa Programme under the grants SEV-2012-0249, FPA2012-34694, and the projects AYA2014-60641-C2-1-P and AYA2012-31101. 
GY acknowledges financial support from MINECO/FEDER (Spain) under project number AYA2012-31101 and AYA2015-63810-P.
The CosmoSim database used in this paper is a service by the Leibniz-Institute for Astrophysics Potsdam (AIP).
The MultiDark database was developed in cooperation with the Spanish MultiDark Consolider Project CSD2009-00064.
The authors gratefully acknowledge the Gauss Center for Supercomputing e.V. (www.gauss-centre.eu) and the Partnership for Advanced Supercomputing in Europe (PRACE, www.prace-ri.eu) for funding the MultiDark simulation project by providing computing time on the GCS Supercomputer SuperMUC at Leibniz Supercomputing Centre (LRZ, www.lrz.de).

\appendix

\section{$V_{max}$ function, measurements and model}
\label{sec:vmax:fun}

The peak circular velocity was proven more efficient than the halo mass to map galaxies to halos \citep{Reddick2013,RodriguezTorres2015,Guo2016}. 
The peak circular velocity ($V_{max}$) is less affected than mass by tidal forces and it thus better defined than halo mass. It traces best the assembly history of the halo and its potential well \citep{Diemand2007b}. 
Thus exists an interest in formulating the halo model in terms of peak velocities instead of mass to obtain more accurate predictions with an analytical model. This section is aimed for a practical use in future exploration of the accuracy of the SHAM/HOD.

Using similar estimators as for the mass function, we measure the velocity function. Figs. \ref{Fig_vmaxfunz1:a}, \ref{Fig_vmaxfunz1:b}  shows the differential velocity function for distinct and satellite subhalos at redshifts below 2.3. We use jackknife as a proxy for errors to perform the fits. 
The analysis of errors is not as careful as previously as we only pretend to provide fitting functions. 
The limits imposed on the $V_{max}$ range are M04, [125, 450]; M10, [250, 800]; M25 and M25n, [600, 1100]; M40 and M40n, [900, 1400] km s$^{-1}$. We estimate a dimension-less velocity function, $V^3/H^3(z)\; dn/dlnV$, the left part of equation (\ref{vf:eqn:model}). As in \citet{Rodriguez-Puebla2016}, we model the measurements as the product of a power law and an exponential cut-off using four parameters 
\begin{eqnarray}
& \log_{10} \left[\frac{V^3}{H^3(z)}\; \frac{dn}{dlnV}\right] (V, A, V_{cut}, \alpha, \beta ) = \cdots \\ 
& \cdots \; \log_{10}\left( 10^A  \left(1+\frac{10^V}{10^{V_{cut}}}\right)^{-\beta} \exp \left[ \left(\frac{10^V}{10^{V_{cut}}}\right)^{\alpha}\right] \right),
\label{vf:eqn:model}
\end{eqnarray}

where $A$ is the normalization, $V_{cut}$ is the cut-off velocity, $\alpha$ the width of the cut-off and $\beta$ the power-law index. We model the redshift trends using an expansion with redshift of each parameter, $p(z)=p_0 + p_1 z + p_2 z ^2 + p_3 z^3 \cdots$.

We fit first the parameters at redshift 0. Then we fit their redshift trends in the range $0\leq z\leq 1$ and then in the range $1\leq z\leq 2.3$. A model with 4 parameters is sufficient at redshift 0. 6 parameters are used to describe the data in each further redshift ranges.
At redshift 0, the fits converge with a reduced $\chi^2=1.43$ for the distinct halos and $\chi^2=0.2$ for the subhalos; see Fig. \ref{Fig:vmax:fun:residual:z0} that shows the residuals of the redshift 0 fits in greater details. Table \ref{table:LFs:fits} gives the parameters of the fits for both populations.

In the range redshift $0\leq z \leq1$, a linear evolution of the parameters $A$ and $V_{cut}$ is sufficient for the fits to converge with a reduced $\chi^2=1.56$ (0.54) for the distinct (satellite); see Fig. \ref{Fig_vmaxfunz1:a} (\ref{Fig_vmaxfunz1:b}) left column row of panels that shows the data, the model and the residuals (from left to right). The parameters $A$ and $V_{cut}$ are compatible in the three redshift bins. Whereas the parameters $\alpha$ and $\beta$ are not. If we add an evolution term for $\alpha$ and $\beta$, the fits converge very slowly and the error on these parameters become very large i.e. current data does not allow to constrain all the parameters at once. Among the parameters, $V_{cut}$ and $A$ are best constrained.

\begin{table}
\caption{Results of model fitting to the $V_{max}$ differential function. Errors are the $1\sigma$ errors. Empty cells mean the parameter was not fitted.}
\label{table:LFs:fits}
\begin{center}
\begin{tabular}{c c c c c}
\hline 
\hline 
\multicolumn{4}{c}{Distinct halos} \\
\hline
z &  &$p^0$ & $p^1$  \\ \hline
$0$ & $A$ & $-0.74\pm0.04$ & \\
 & $V_{cut}$ & $2.94\pm0.02$ & \\
 & $\alpha$  & $2.02\pm0.08$ & \\
 & $\beta$  & $-0.79\pm0.24$ & \\
 & $\chi^2$ & $286.11/199=1.43$ \\
\hline
$0\leq z\leq 1$ & $A$ & $-0.71\pm0.08$ & $-0.62\pm0.03$ \\
 & $V_{cut}$ & $2.93\pm0.09$ & $-0.176\pm0.001$ \\
 & $\alpha$  & $1.782\pm0.07$ &  \\
 & $\beta$  & $-0.82\pm0.07$ & \\
 & $\chi^2$ & \multicolumn{2}{c}{$2504.8/1599=1.56$} \\
\hline
$1\leq z\leq 2.3$ & $A$ & $-0.71\pm0.14$ & $-0.62\pm0.05$ \\
 & $V_{cut}$ & $2.85\pm0.07$ & $-0.15\pm0.02$ \\
 & $\alpha$  & $1.58\pm0.77$ &  \\
 & $\beta$  & $-0.77\pm0.02$ & \\
 & $\chi^2$ & \multicolumn{2}{c}{$1555.6/1039=1.49$} \\
\hline 
\hline 
\multicolumn{4}{c}{Satellite halos} \\
\hline
z &  &$p^0$ & $p^1$  \\ \hline
$0$ & $A$ & $-1.66\pm0.01$ & \\
 & $V_{cut}$ & $2.69\pm0.01$ & \\
 & $\alpha$  & $1.57\pm0.02$ & \\
 & $\beta$  & $0.36\pm0.02$ & \\
 & $\chi^2$ & $37.6/185=0.20$ \\
\hline
$0\leq z\leq 1$ & $A$ & $-1.67\pm0.07$ & $-0.62\pm0.08$ \\
 & $V_{cut}$ & $2.71\pm0.05$ & $-0.14\pm1.$ \\
 & $\alpha$  & $1.626\pm0.08$ &  \\
 & $\beta$  & $-0.48\pm0.01$ & \\
 & $\chi^2$ & \multicolumn{2}{c}{$591.8/1081=0.54$} \\
\hline
$1\leq z\leq 2.3$ & $A$ & $-1.45\pm0.08$ & $-0.63\pm0.05$ \\
 & $V_{cut}$ & $2.53\pm0.05$ & $-0.14\pm0.03$ \\
 & $\alpha$  & $1.23\pm0.12$ &  \\
 & $\beta$  & $0.03\pm0.11$ & \\
 & $\chi^2$ & \multicolumn{2}{c}{$274.0/470=0.58$} \\
\hline
\end{tabular}
\end{center}
\end{table}

\begin{figure*}
\begin{center}
\includegraphics[type=png,ext=.png,read=.png,width=7.5cm]{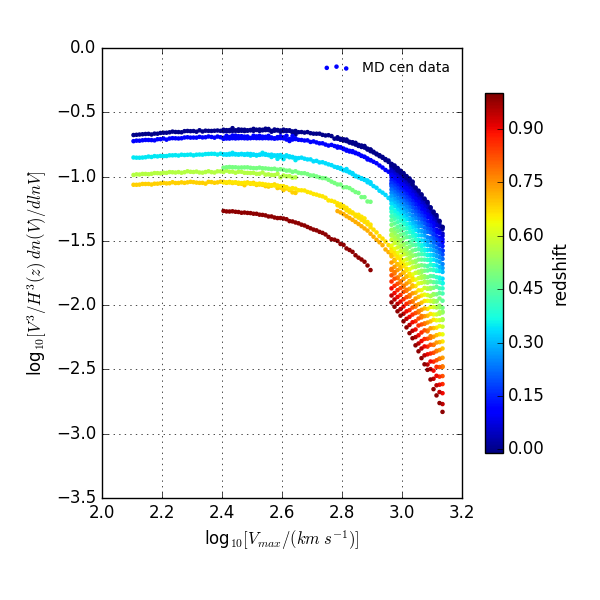} 
\includegraphics[type=png,ext=.png,read=.png,width=7.5cm]{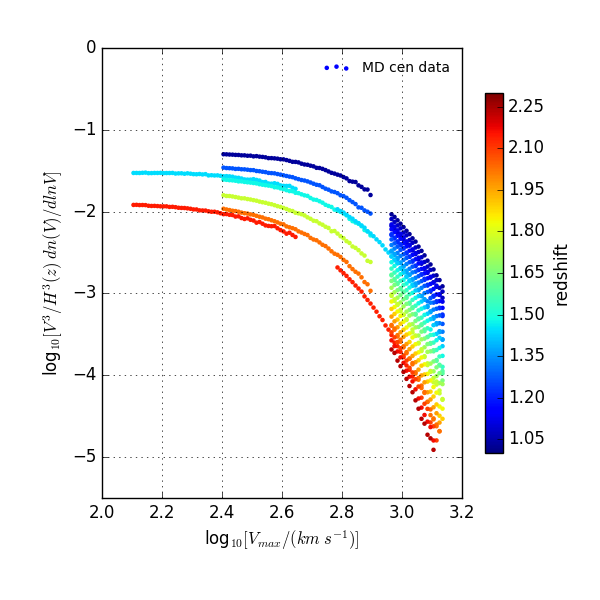}

\includegraphics[type=png,ext=.png,read=.png,width=7.5cm]{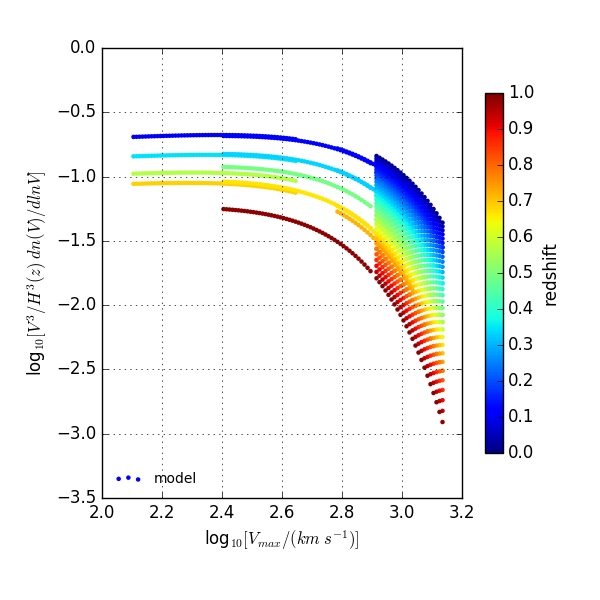} 
\includegraphics[type=png,ext=.png,read=.png,width=7.5cm]{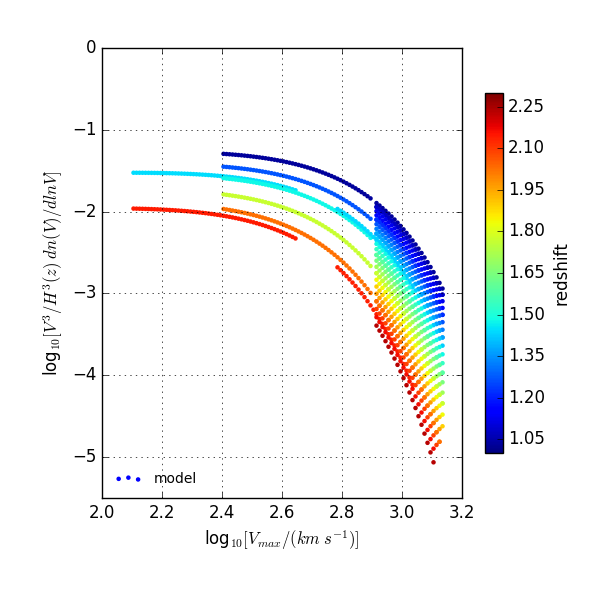}

\includegraphics[type=png,ext=.png,read=.png,width=7.5cm]{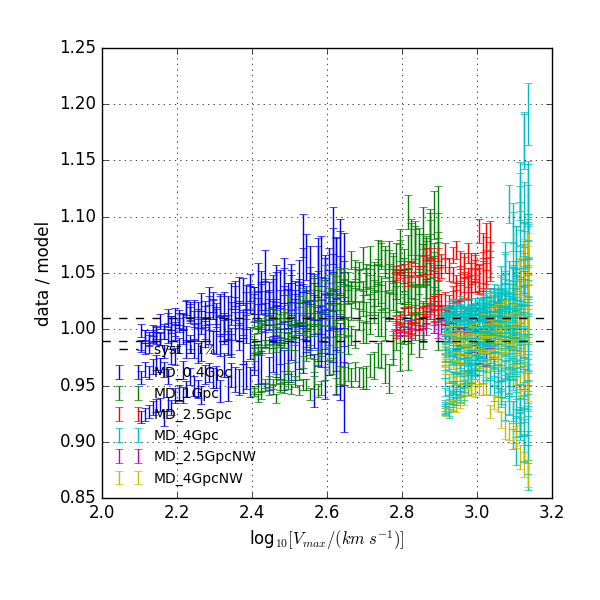} \includegraphics[type=png,ext=.png,read=.png,width=7.5cm]{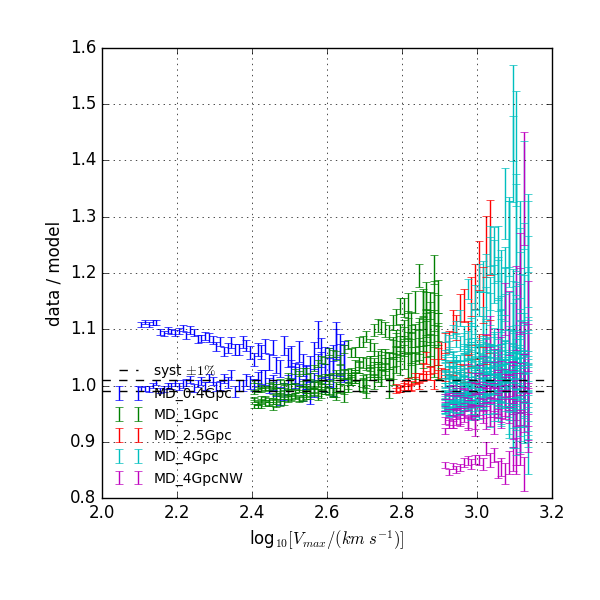}
\caption{Measurements of the differential distinct halo $V_{max}$ function vs. $V_{max}$ colored with redshift (top row), its model (middle) and the residuals around the model (bottom row). The first column shows the range $0\leq z \leq 1$ and the second column the $1\leq z \leq 2.3$ range. Residual around the $0\leq z \leq 1$ model are contained in $\pm15\%$ and $\pm20\%$ for the high redshift range.}
\label{Fig_vmaxfunz1:a}
\end{center}
\end{figure*}

\begin{figure*}
\begin{center}
\includegraphics[type=png,ext=.png,read=.png,width=7.5cm]{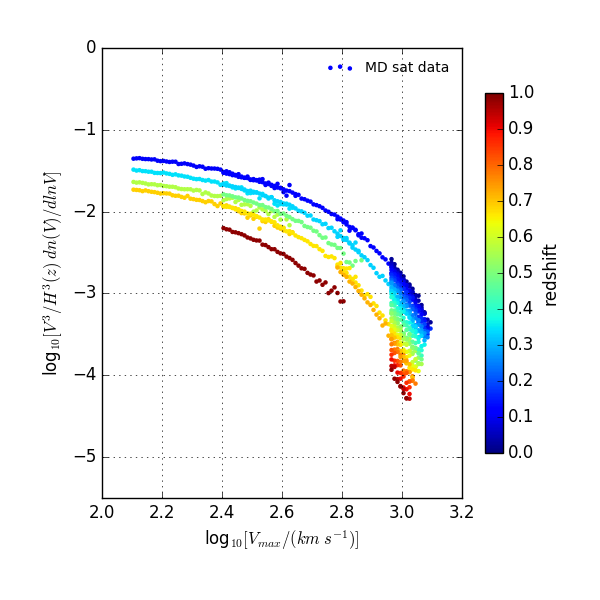} 
\includegraphics[type=png,ext=.png,read=.png,width=7.5cm]{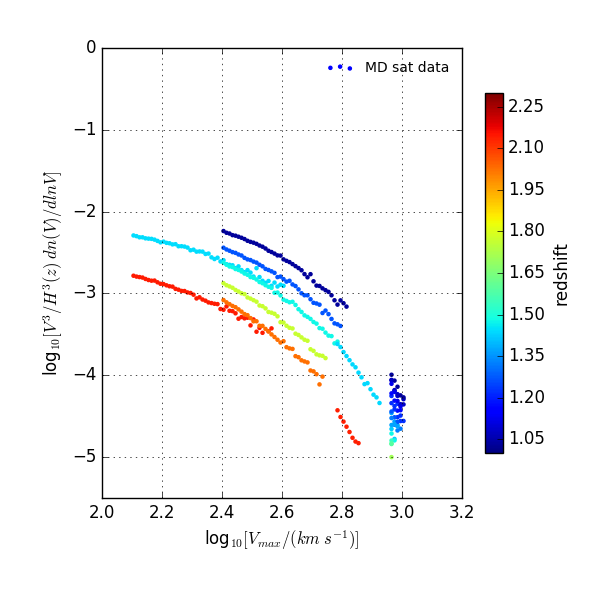}

\includegraphics[type=png,ext=.png,read=.png,width=7.5cm]{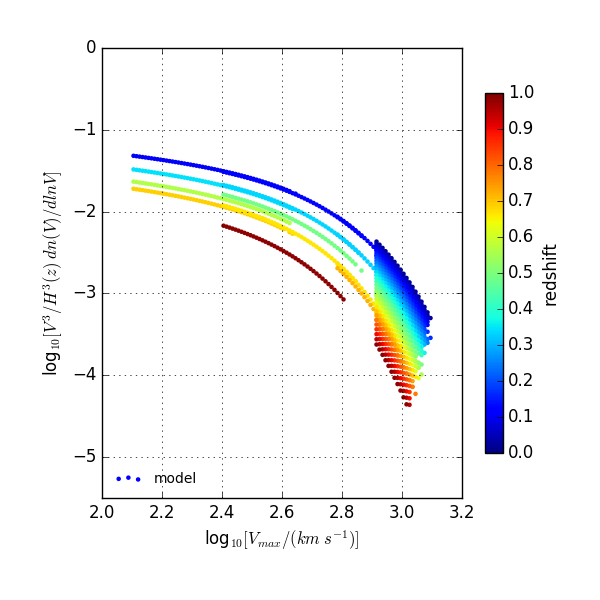} 
\includegraphics[type=png,ext=.png,read=.png,width=7.5cm]{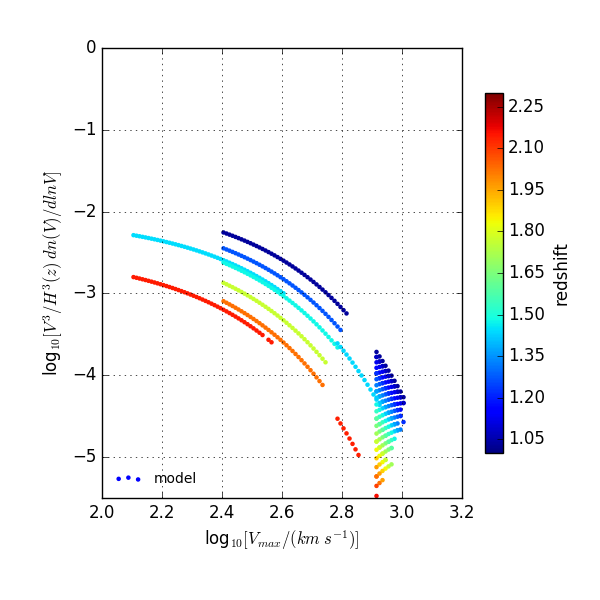}

\includegraphics[type=png,ext=.png,read=.png,width=7.5cm]{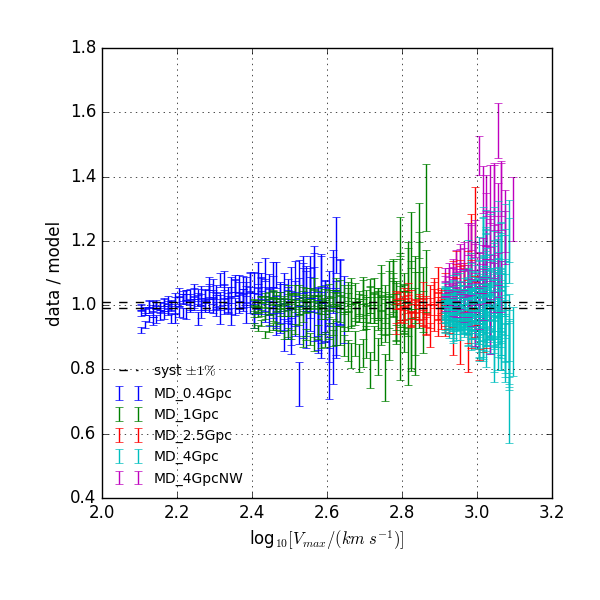} 
\includegraphics[type=png,ext=.png,read=.png,width=7.5cm]{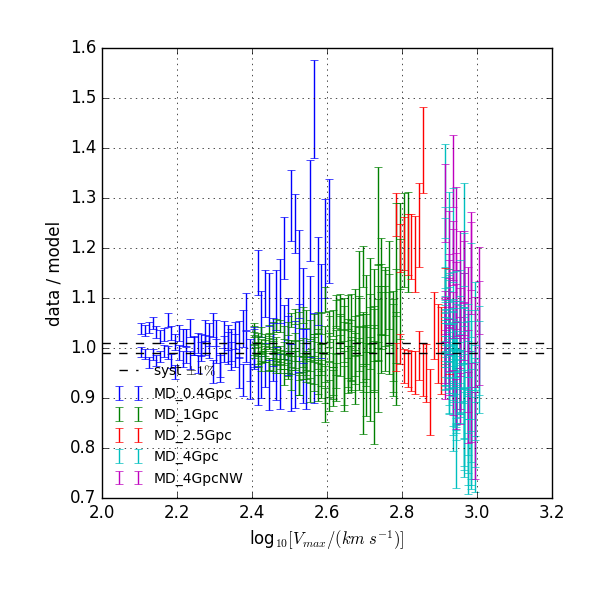}
\caption{Continued Fig. \ref{Fig_vmaxfunz1:a} for the satellite subhalos in the same redshift ranges. Residuals are of the same order of magnitude as for the distinct halos.}
\label{Fig_vmaxfunz1:b}
\end{center}
\end{figure*}

\begin{figure*}
\begin{center}
\includegraphics[type=png,ext=.png,read=.png,width=8cm]{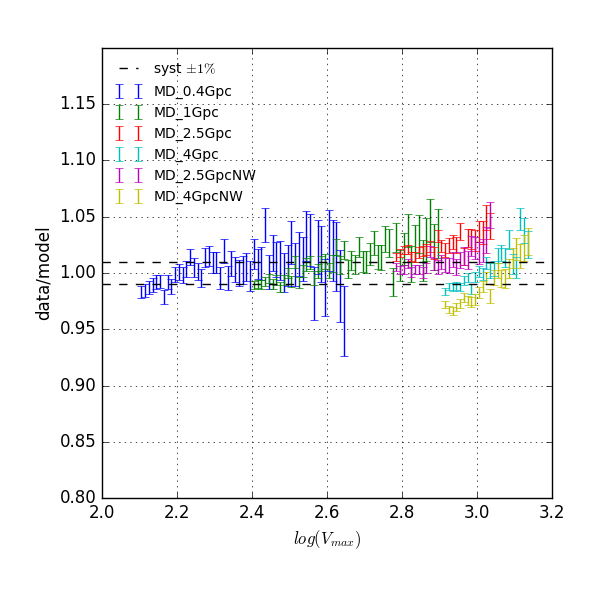} 
\includegraphics[type=png,ext=.png,read=.png,width=8cm]{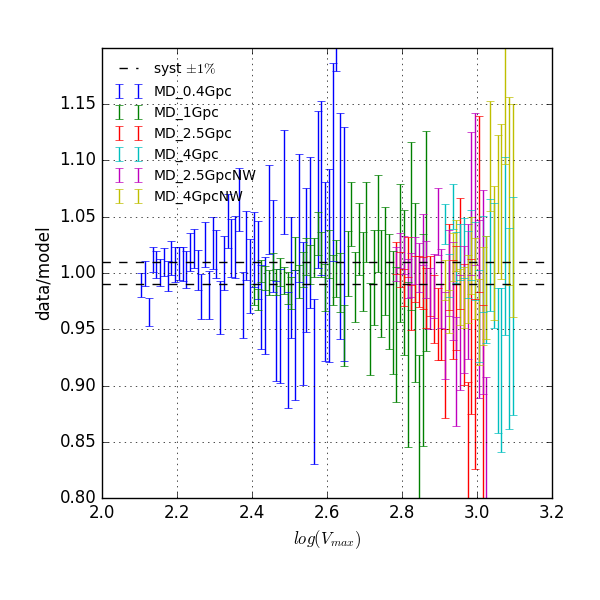} 
\caption{Residuals around the redshift 0 model are well $\pm5\%$ for the distinct halos (left) and within $\pm10\%$ for the satellite halos (right).}
\label{Fig:vmax:fun:residual:z0}
\end{center}
\end{figure*}

\end{document}